\documentclass[english]{article}

\usepackage[T1]{fontenc}
\usepackage[latin9]{inputenc}

\usepackage[a4paper]{geometry}
\usepackage{setspace}
\usepackage{amsmath}
\usepackage{amssymb}

\usepackage{graphicx}

\usepackage{cite}

\usepackage{hyperref}

\DeclareMathOperator{\rmd}{d}

\title{Understanding the nature of the long--range memory phenomenon in socioeconomic systems}
\author{Rytis Kazakevicius, Aleksejus Kononovicius, Bronislovas Kaulakys, Vygintas Gontis}
\date{\small Institute of Theoretical Physics and Astronomy, Vilnius University, Sauletekio al. 3, 10257 Vilnius, Lithuania}

\begin{document}

\maketitle

\begin{abstract}
In the face of the upcoming 30th anniversary of econophysics, we review
our contributions and other related works on the modeling of
the long--range memory
phenomenon in physical, economic, and other social complex systems. Our group
has shown that the long--range memory phenomenon can be reproduced
using various Markov processes, such as point processes, stochastic
differential equations and agent--based models. Reproduced
well enough to match other statistical properties of the financial
markets, such as return and trading activity distributions and first--passage
time distributions. Research has lead us to question whether the
observed long--range memory is a result of actual long--range memory
process or just a consequence of non--linearity of Markov processes.
As our most recent result we discuss the long--range memory of the
order flow data in
the financial markets and other social systems from the perspective of the
fractional L\`{e}vy stable motion. We test widely used long--range memory estimators on discrete fractional L\`{e}vy stable motion represented by the ARFIMA sample series.
Our newly obtained results seem indicate that new estimators of
self--similarity and long--range memory for analyzing systems with
non--Gaussian distributions have to be developed.
\end{abstract}

\section{Introduction}

Many empirical data sets and theoretical models
have been investigated using the tool of spectral analysis. Many researchers across different
fields find the power spectral density (abbr. PSD) of the $1/f^{\beta}$
form (with $0.5 \lesssim \beta \lesssim 1.5$) to be of a particular interest
\cite{Mandelbrot1968SIAMR,Press1978CA,Dutta1981RMP,Bak1987,West1989IJMPB,Mandelbrot-1999,Milotti2002,Ward2007Scholarpedia,Rodriguez2014PRE,Yadav2021}.
Both because of its apparent omnipresence and the implication of slowly decaying
autocorrelation, which indicates the presence of the long--range memory
phenomenon. Long--range memory
is also one of the established stylized facts of the financial markets
\cite{Taqqu1995Fractals,Gopikrishnan1998EPJB,Plerou2000PRE,Cont2001RQUF,Mantegna2000Cambridge,Gabaix2003Nature,Farmer2004QF,Gabaix2006QJE,Alfi2009EPJB1}. Consequently, as our group was
investigating $1/f$ noise
\cite{Kaulakys1995NPSF,Kaulakys1998PRE,Kaulakys1999PLA,Kaulakys2005PhysRevE}, we have become naturally interested in
the rapidly growing field of econophysics.
The term ``econophysics'' being coined by H.~E.~Stanley
in Statphys conference in Kolkata in 1995 \cite{Ghosh2013KSMR}.
Over these last three decades econophysics has matured both from the
theoretical and the applied perspectives. Here we review mostly our own
and directly adjacent approaches, therefore we would like to recommend a
couple of broader reviews, which can be found in
\cite{Pereira2017PhysA,Jovanovic2017OUP}.

Our first publications were devoted to the modeling of the
financial markets
\cite{Gontis2001LietFizRink,Gontis2002MC}. In those works we have considered
trades occurring in the financial markets as point events driven by a point
process proposed in
\cite{Kaulakys1998PRE,Kaulakys1999PLA,Kaulakys2005PhysRevE}. Thanks to
the organizers of the international conference Applications of Physics in
Financial Analysis 4, held in Warsaw in 2003, we were able to
present our findings to econophysicists. Our first results inspired by
the interaction with the participants of the APFA4 conference have
been published in \cite{Gontis2004PhysA343,Gontis2004PhysA344}. We presented
our ideas in the more general context of complex systems in
\cite{Gontis2004SoldStPhen,Gontis2005AIP}.

Later, we have taken part in COST Action P10 ``Physics of Risk'' and
the follow--up COST Action MP0801 ``Physics of Competition and Conflicts''.
Bronislovas Kaulakys and Vygintas Gontis were executive committee
members of both COST Actions, while the other group members
gave talks and poster presentations during the annual meetings and helped organize an annual Action meeting in Vilnius in 2006. This COST Action has helped us embrace econophysics and be recognized as econophysicists.

While it may be natural to see trades in the financial markets as
point events
\cite{Gontis2001LietFizRink,Gontis2002MC,Gontis2004PhysA343,Gontis2004PhysA344},
modeling volatility and return as a point process was not as
straightforward. We have developed our approach further by abstracting the
point process away and considering a continuous framework of Langevin
stochastic differential equations (abbr. SDEs). First we have shown that the
continuous interpretation of the point process model
works well for trading activity \cite{Gontis2007PhysA}. And thus we have
refined the SDE approach with model for volatility and return \cite{Gontis2006JStatMech,
Gontis2010Intech,Gontis2010PhysA,Gontis2011JDySES,Ruseckas2012ACS}.
Interestingly similar SDEs can be derived from a simple agent--based model
(abbr. ABM) \cite{Ruseckas2011EPL,Kononovicius2012PhysA}, too. With time
we have developed more
complicated ABMs to account for the separation of time scales and
order flow \cite{Gontis2014PlosOne,Kononovicius2019OB}. We even have
branched out into sociophysics
\cite{Kononovicius2017Complexity,Kononovicius2019CompJStat,Kononovicius2020JStatMech,Kononovicius2021CSF}
as we have understood that the herding ABM we used to model the financial
market is essentially equivalent to the well--known voter model
\cite{Castellano2009RevModPhys,Dong2018IntFus,Noorazar2020EPJP}.

For $10$ months (in 2015 and 2016), Vygintas Gontis, with the support of the Baltic American Freedom Foundation
has stayed as a visiting researcher at the Center of Polymer Studies of Boston University. Discussions with the founding fathers of econophysics H.~E.~Stanley, professors Sh.~Havlin, B.~Podobnik, and S.~Buldyrev,
resulted in a paper \cite{Gontis2016PhysA}.
Together we have considered volatility return intervals (term
inspired by the studies \cite{Yamasaki2005PNAS,Wang2006PhysRevE,
Wang2008PhysRevE,Denys2016PhyRevE}) of the financial time series at various time scales. In
the paper we have shown that the time intervals between large financial
fluctuations is distributed according to a power--law probability density
function (abbr. PDF) $p\left(\tau\right)\sim\tau^{-3/2}$
\cite{Gontis2016PhysA}.
The same distribution arise in our models and from many other
one--dimensional Markov processes \cite{Redner2001Cambridge},
while the long--range memory process would exhibit a different distribution,
such as $p\left(\tau\right)\sim\tau^{2-H}$, which is a well--known
result for the fractional Brownian motion (abbr. FBM) \cite{Ding1995fbm}.

Here we provide an overview of our approach to understanding and modeling
the long--range memory phenomenon in financial markets and other
complex systems and share our most recent result. In Section~\ref{sec:pp-sde}
we introduce the original point process and
discuss how to derive a non--linear SDE, which can reproduce
the long--range memory phenomenon. We also discuss numerous extensions of
both the point process model and  non--linear SDE. Next, Section~\ref{sec:abm},
we show how we can obtain a similar SDE from a simple herding ABM.
Following the overview, we also present a novel result,
which concerns understanding the nature of the self--similarity and long--range memory
phenomenon from the perspective of fractional L\`{e}vy stable motion
(abbr. FLSM)
and auto--regressive fractionally integrated moving average
(abbr. ARFIMA) time series. In Section~\ref{sec:arfima}
we tested various long--range memory estimators such as Mean squared
displacement, method of Absolute Value estimator, Higuchi's method, and
burst and interburst duration analysis on  fractional Levy stable motion
(ARFIMA(0,d,0) time series). Finally, in
Section~\ref{sec:future-considerations}, we share our future considerations.

\section{The multiplicative point process, the class of stochastic differential
equations and their applications\label{sec:pp-sde}}

In this section, we overview how the physically motivated point process
proposed in \cite{Kaulakys1998PRE,Kaulakys1999PLA,Kaulakys2005PhysRevE} was
applied to model trading activity and absolute returns in the financial
markets. We also discuss numerous extensions of the model into some related
research topics, such as superstatistics, anomalous and non--homogeneous
diffusion.

\subsection{The multiplicative point process model}

Let us consider signal $I\left(t\right)$ composed of pulses with
profiles given by $A_{k}\left(x\right)$:
\begin{equation}
I\left(t\right)=\sum\limits _{k}A_{k}\left(t-t_{k}\right),\label{eq:pp-signal}
\end{equation}
here $t_{k}$ is the event (pulse) time. There are many physical and
social systems, which generate signals of such nature: electric current
\cite{Johnson1925PR}, music \cite{Levitin2012PNAS}, human heartbeat
\cite{Kobayashi1982BioMed}, internet traffic \cite{Gontis2005AIP}
or trading activity \cite{Gontis2004PhysA343} to name a few.

As most profiles of the pulses are brief, it is trivial that they
would influence only high frequencies corresponding to the typical inverse
pulse length. If we are interested in longer--term dynamics it is sufficient
to assume that the Kronecker delta function well approximates the profile, $A_{k}\left(x\right)=a_{k}\delta\left(x\right)$. Many such
systems are driven by the flow of identical or similar objects, such
as electrons, packets, or trades. This lets us simplify \eqref{eq:pp-signal}
and investigate it as a temporal point process with unit events. Such
process can be either described by the event times $\left\{ t_{k}\right\} $
or by the inter--event times $\left\{ \tau_{k}=t_{k+1}-t_{k}\right\} $.

The inter--event times are far more convenient choice to model as they
at least can give a semblance of the stationarity, while event times
are obviously non--stationary as $\left\{ t_{k}\right\} $ is monotonically
increasing series. In
\cite{Kaulakys1998PRE,Kaulakys1999PLA,Kaulakys2005PhysRevE} it was analytically
shown that a relatively slow autoregresive AR(1) Brownian motion of $\tau_{k}$ yield
$1/f$ fluctuations of the signal $I(t)$. \cite{Gontis2004PhysA343}
has built upon this observation and introduced multiplicative point
process for the inter--event time
\begin{equation}
\tau_{k+1}=\tau_{k}+\sigma^{2}\gamma\tau_{k}^{2\mu-1}+\sigma\tau_{k}^{\mu}\varepsilon_{k}.\label{eq:pp-tau}
\end{equation}
In the above, it is assumed that inter--event time fluctuates due
to exogenous perturbations. Perturbations are assumed to be standard
uncorrelated Gaussian random variables, $\varepsilon_{k}$. The general
rate of change is governed by $\sigma$, while $\gamma$ is the damping
constant. Multiplicativity, specified by $\mu$, ensures that $I\left(t\right)$
is multifractal and has a power--law PDF.
This point process model has found its use for the analysis of
$1/f$ noise and long--range memory in many diverse phenomena such as musical
rhythm spectra \cite{Levitin2012PNAS}, human cognition
\cite{Wagenmakers2004PsychonBullRev}, human interaction dynamics
\cite{Mathiesen2013PNAS}, turbulence \cite{Leonardis2012AstrophysJ} and few
others \cite{Meskauskas2005AIP,Ribeiro_2017,Ribeiro_2018,Nakamura2019PRE}.
Inspired by this model \cite{Erland_2007} has shown under which conditions
$1/f^\beta$ spectrum can arise from reversible Markov chains.

After closer examination it should be evident that Eq.~\eqref{eq:pp-tau}
can be seen as an iterative solution of a certain SDE if Euler--Maruyama
method was used \cite{Kloeden1999Springer}. Hence the corresponding Langevin
SDE can be trivially recovered from the iterative relation \eqref{eq:pp-tau}:
\begin{equation}
\rmd\tau =\sigma^{2}\gamma\tau^{2\mu-1}\rmd k+\sigma\tau^{\mu}\rmd W_{k}.\label{eq:tau-k-sde}
\end{equation}
Here $W_k$ is standard uncorrelated Wiener process in $k$--space (hence the
index). It is important to note that this SDE is in the $k$--space (or event
space) and not in the real
time. Also this SDE must be solved by restricting the diffusion of the
inter--event time $\tau$ to some arbitrary interval
$\left[\tau_{\mathrm{min}},\text{\ensuremath{\tau_{\mathrm{max}}}}\right]$
on the positive half--plane as otherwise this SDE may not have a
stationary distribution. If stationary distribution exists, then the
stationary PDF of $\tau$ is a power--law:
\begin{equation}
p_{k}\left(\tau\right)=\frac{\alpha+1}{\tau_{\mathrm{max}}^{\alpha+1}-\tau_{\mathrm{min}}^{\alpha+1}}\tau^{\alpha},\quad\alpha=2\left(\gamma-\mu\right).
\end{equation}

\begin{figure}
\centering
\includegraphics[width=0.9\textwidth]{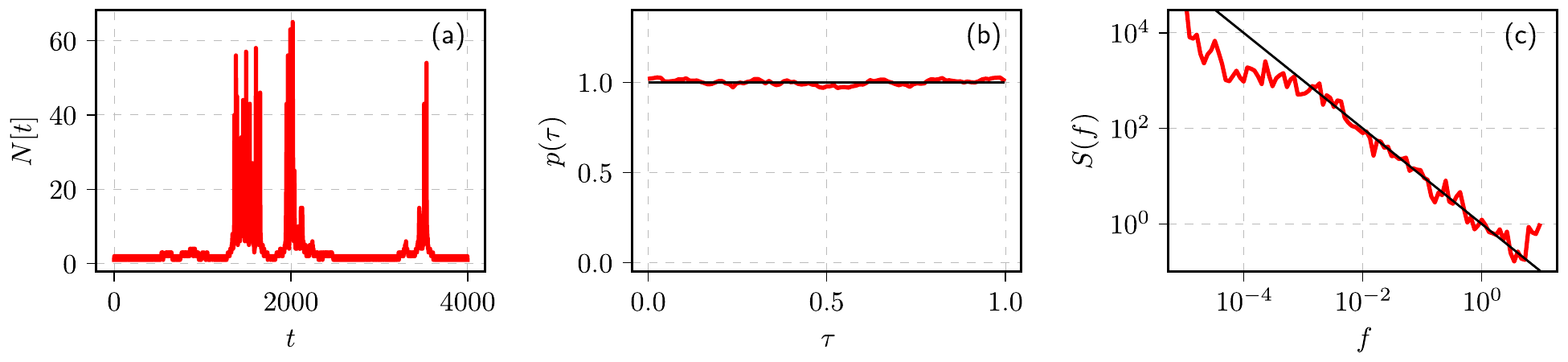}
\caption{Statistical properties of the point process by numerically solving
Eq.~\eqref{eq:pp-tau}: (a) sample fragment of corresponding $N\left[t\right]$
time series, (b) PDF of the inter--event times and (c) PSD of the
process. Red curves correspond to numerical results, while black curves
are theoretical power--law fits with (b) $\alpha=0$ and (c) $\beta=1$.
Model parameter values: $\gamma=0$, $\mu=0$, $\sigma=0.1$, $w=1$.
\label{fig:pp-stats}}
\end{figure}

Yet the main result of \cite{Gontis2004PhysA343} is the power--law
statistical properties of $I\left(t\right)$. In the limit $\tau_{\mathrm{min}}\rightarrow0$
and $\tau_{\mathrm{max}}\rightarrow\infty$ PSD of $I\left(t\right)$
in arbitrarily long range of frequencies has a power--law slope:
\begin{equation}
S\left(f\right)\sim1/f^{\beta},\quad\beta=1+\frac{2\left(\gamma-\mu\right)}{3-2\mu}.
\end{equation}
Number of events in a selected time window, for example number of
trades per minute, also has a power--law distribution \cite{Gontis2004PhysA343}:
\begin{equation}
p\left(N\right)\sim N^{-2\left(\gamma-\mu\right)-3}.
\end{equation}
Formally one could define the number of events in a window of length
$w$ as $N\left[t\right]=\int_{t}^{t+w}I(u)\rmd u$ (here the square
brackets indicate that $N$ is in discrete time). These analytical
results can be confirmed by numerical simulation (see Fig.~\ref{fig:pp-stats}).

\subsection{The class of non--linear stochastic differential equations}

In \cite{Kaulakys2004PhysRevE,Kaulakys2006PhysA,Gontis2007PhysA,Kaulakys2009JStatMech} we have made a transition
from $k$--space
to real time and this enabled us to model trading activity and absolute
returns in the financial markets not only qualitatively, but quantitatively,
too. The transition from SDE in $k$--space, Eq.~\eqref{eq:tau-k-sde},
to real time is done by substitution $\rmd t=\tau\rmd k$, which yields:
\begin{equation}
\rmd\tau=\sigma^{2}\gamma\tau^{2\mu-2}\rmd t+\sigma\tau^{\mu-1/2}\rmd W.
\end{equation}
Here $W$ is standard uncorrelated Wiener process.
Modeling inter--event time in real time makes less sense than in the $k$--space,
so let us change the variable to the number of events per unit time
$x=\frac{1}{\tau}$. Applying It\^o transformation yields:
\begin{equation}
\rmd x=\sigma^{2}\left(\eta-\frac{\lambda}{2}\right)x^{2\eta-1}\rmd t+\sigma x^{\eta}\rmd W.\label{eq:sde-class}
\end{equation}
In the above we have introduced a more convenient set of parameters:
\begin{equation}
\eta=\frac{5}{2}-\mu,\quad\lambda=2\left(\gamma-\mu\right)+3.
\end{equation}

\begin{figure}
\centering
\includegraphics[width=0.7\textwidth]{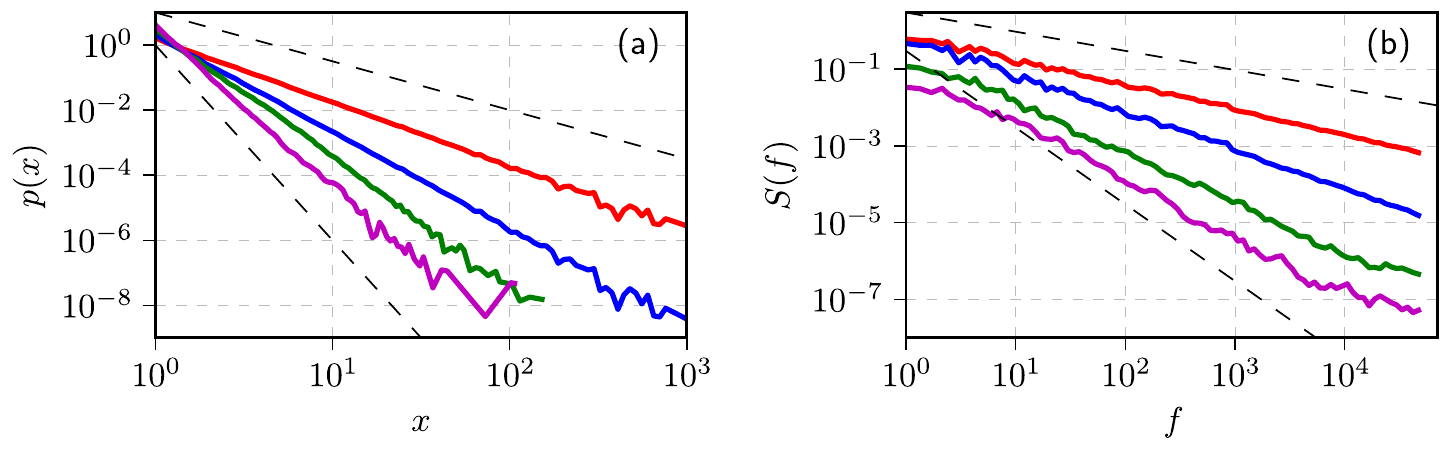}
\caption{Various slopes of PDF (a) and PSD (b) reproduced by the numerical
solutions of SDE~\eqref{eq:sde-class}. Model parameter values: $\sigma=1$,
$\eta=2.5$ (all cases) and $\lambda=2$ (red curves in both (a) and
(b)), $3$ (blue curves), $4$ (green curves) and $5$ (magenta curves).
Black dashed lines correspond to (a) $p\left(x\right)\sim x^{-\lambda}$
with $\lambda=1.5$ and $\lambda=6$ (upper and lower curves), (b)
$S\left(f\right)\sim1/f^{\beta}$ with $\beta=0.5$ and $\beta=2$
(upper and lower curves).\label{fig:sde-stats}}
\end{figure}

As far as SDE~\eqref{eq:sde-class} corresponds to the point process
defined by Eq.~\eqref{eq:pp-tau}, the results for stationary PDF
and PSD should apply:
\begin{align}
p\left(x\right) & \sim x^{-\lambda},\qquad S\left(f\right)\sim1/f^{\beta},\quad\beta=1+\frac{\lambda-3}{2\eta-2}.\label{eq:sde-class-stats}
\end{align}
The validity of these theoretical predictions was extensively checked
numerically
(see Fig.~\ref{fig:sde-stats} for a quick example) and also, in
\cite{Ruseckas2010PhysRevE}, proven analytically. Analytical proof
provided in \cite{Ruseckas2010PhysRevE} allows interpreting the process
modeled by SDE~\eqref{eq:sde-class} in a more general context. In
fact we can model any process possessing these power--law statistical
properties, even processes, which make less sense from the perspective
of the original point process.

Eq.~\eqref{eq:sde-class} and similar random walk models have
been used to model the EUR/CHF exchange rate \cite{Lera2015}. It has also
lead to numerous modifications by our group, which we discuss in detail in
the following subsections.

\subsection{Reproducing the long--range memory using GARCH(1,1) process}

Autoregressive conditional heteroscedasticity (abbr. ARCH) family models
\cite{Engle1982Econometrica,Bollerslev1986Econometrics,Engle1986EcoRev,Potters1998EPL,Giraitis2000AAP,Bollerslev2008CREATES}
are quite popular forecasting tools among professional traders as
well as researchers interested in the long--range memory phenomenon.
Unlike SDEs ARCH family models have explicitly built--in memory.
Which is built--in either via explicit dependence on the numerous
previous states, infinitely many in case of ARCH($\infty$) model
\cite{Giraitis2007,Giraitis2009,Giraitis2018ET}, or via fractional integration procedure,
which introduces memory similar to the one present in the fractional
Brownian motion, as
in fractionally integrated GARCH (abbr. FIGARCH) model \cite{Granger1980JTSA,Baillie1996JE,Tayefi2012}.
In \cite{Kononovicius2015PhysA} we have shown that it is possible
to modify GARCH(1,1) model, which is Markovian in nature, to reproduce
$1/f$ spectrum.

Generalized autoregressive conditional heteroskedasticity (abbr. GARCH) processes can be approximated by the diffusion processes. There
are two competing approaches, which yield continuous approximations
of GARCH processes using sets of SDEs. One of the approaches was proposed
by Nelson \cite{Nelson1990JEco} and the other by Kluppelberg \textit{et
al.} \cite{Kluppelberg2004JApplProbab,Kluppelberg2010SSRN}. In case
of GARCH(1,1) Nelson's approach is easier to apply, but has a drawback
that the resulting COGARCH(1,1) would be driven by two source of noise,
instead of the one in the GARCH(1,1). Yet we can circumvent the problem
by ignoring the observed heteroskedastic economic variable $z_{t}$
and focusing on the approximation of the volatility process, $\sigma_{t}^{2}$,
of GARCH(1,1):
\begin{align}
z_{t} & =\sigma_{t}\omega_{t},\\
\sigma_{t}^{2} & =a+bz_{t-1}^{2}+c\sigma_{t-1}^{2}=a+b\sigma_{t-1}^{2}\omega_{t-1}^{2}+c\sigma_{t-1}^{2}.\label{eq:garch-volatility}
\end{align}
In the above $\omega_{t}$ is the noise, while $a$, $b$ and $c$
are the GARCH(1,1) model parameters. For Nelson's approach to work
we need to compute first and second moments of change in volatility.
With the usual GARCH(1,1) we obtain SDE for geometric Brownian motion
\cite{Kononovicius2015PhysA}.

Now lets introduce non--linearity into Eq.~\eqref{eq:garch-volatility}.
In \cite{Kononovicius2015PhysA} we have explored two such options:
\begin{align}
\sigma_{t}^{2} & =a+b\sigma_{t-1}^{\mu}\omega_{t-1}^{\mu}+c\sigma_{t-1}^{2},\\
\sigma_{t}^{2} & =a+b\sigma_{t-1}^{\mu}\left|\omega_{t-1}\right|^{\mu}+\sigma_{t-1}^{2}-c\sigma_{t-1}^{\mu}.
\end{align}
Both of these options can be approximated by SDEs belonging to the
class of SDEs~\eqref{eq:sde-class} with $\lambda=\mu$ and $\eta=\mu/2$.
Consequently both of these options reproduce $1/f$ spectrum with
$\mu=3$. Other parameters,
$a$, $b$ and $c$, influence only the additional terms, which restrict
the diffusion of $\sigma_{t}^{2}$. Setting these values too high
shrinks the interval and the power--law distribution becomes extremely
hard to observe.

\subsection{Anomalous diffusion in the long--range memory process}

SDE~\eqref{eq:sde-class} can be also seen to describe a heterogeneous
diffusion in a non--linear potential. Such diffusion leads to anomalous
growth in variance \cite{Kazakevicius2016PRE}
\begin{equation}
\left\langle \left[ x(t)- \left\langle x(t) \right\rangle \right]^2
\right\rangle \sim t^{\theta}, \quad \theta = \frac{1}{1-\eta} .
\end{equation}
This phenomenon is also known as anomalous diffusion
\cite{Havlin2002AP,Avraham2005CUP,Metzler2014PCCP}.
If $\theta = 1$ then the process exhibits normal diffusion. Otherwise if
$0<\theta<1$, the diffusion is slower than normal and is referred to as
sub--diffusion. The diffusion may also be faster, if $1<\theta<2$, in that
case it is called super--diffusion.

The anomalous diffusion can be obtained from SDE~\eqref{eq:sde-class} only
for specific parameter  values such as $\lambda<1$ and $\eta<1/2$
\cite{Kazakevicius2016PRE}. Because power--law slope of the PSD, $\beta$,
varies between $0$ and $2$, from
Eq.~\eqref{eq:sde-class-stats} follows that anomalous diffusion and
power--law noise can be observed at the same time only for negative
parameter $\eta$
values, specifically for $\eta<(\lambda-1)/2$ and $\lambda<1$. However,
for these
parameters values numerical simulation would become very slow and
inefficient \cite{Ruseckas2010PhysRevE}. Therefore we have considered
generalizing SDE~\eqref{eq:sde-class} by considering non--Gaussian white
noise.

In \cite{Kazakevicius2014PhysA} we have considered L\'evy $\alpha$--stable
noise. SDE equivalent to SDE~\eqref{eq:sde-class}, but with L\'evy
$\alpha$--stable noise takes the following form:
\begin{equation}
\frac{\rmd x}{\rmd t}=\gamma(\eta,\lambda,\alpha)
x^{\alpha(\eta-1)+1}+x^{\eta}\xi_{\alpha}(t).\label{eq:sde-levy}
\end{equation}
Here $\xi_{\alpha}(t)$ is a white noise, intensity of which is
distributed according to the symmetric L\'evy $\alpha$--stable distribution.
Characteristic function of the noise intensity is given by:
\begin{equation}
\left\langle\exp\left(ik\xi_{\alpha}\right)\right\rangle =
\exp\left(-\sigma^{\alpha}\left|k\right|^{\alpha} \right).
\end{equation}
Here $\alpha$ is the index of stability and $\sigma$ is the scale
parameter. We interpret SDE~\eqref{eq:sde-levy} in It\^o sense and it can also
be written in the form
\begin{equation}
\rmd x=\gamma(\eta,\lambda,\alpha) x^{\alpha(\eta-1)+1}\rmd t+x^{\eta}\rmd
L_{t}^{\alpha}.
\end{equation}
Here $\rmd L_{t}^{\alpha}$ stands for the increments of L\'evy $\alpha$--stable
motion $L_{t}^{\alpha}$. If SDE~\eqref{eq:sde-levy} is solved with reflective
boundary conditions and
\begin{equation}
\gamma(\eta,\lambda,\alpha)=\frac{\sin\left[\pi\left(\frac{\alpha}{2}-\alpha\eta+\lambda\right)\right]}{\sin[\pi(\alpha(\eta-1)-\lambda)]}\frac{\Gamma(\alpha\eta-\lambda+1)}{\Gamma(\alpha(\eta-1)-\lambda+2)},
\end{equation}
then generalized SDE~\eqref{eq:sde-levy} generate time series with power--law
steady--state PDF and power--law PSD:
\begin{equation}
p(x)\sim x^{-\lambda}, \qquad S(f)\sim\frac{1}{f^{\beta}},\quad \beta=1+\frac{\lambda-3}{\alpha(\eta-1)}.
\end{equation}
Extensive numerical simulations have shown that due to the presence of
the multiplicative L\'evy $\alpha$--stable noise in Eq.~\eqref{eq:sde-levy} both
sub--diffusion and super--diffusion can be observed together with power--law
noise even for positive $\eta$ values \cite{Kazakevicius2015ChSF}. However,
no analytical expression for anomalous diffusion exponent dependence on SDE
parameters has been derived yet.

\begin{figure}
\centering
\includegraphics[width=0.9\textwidth]{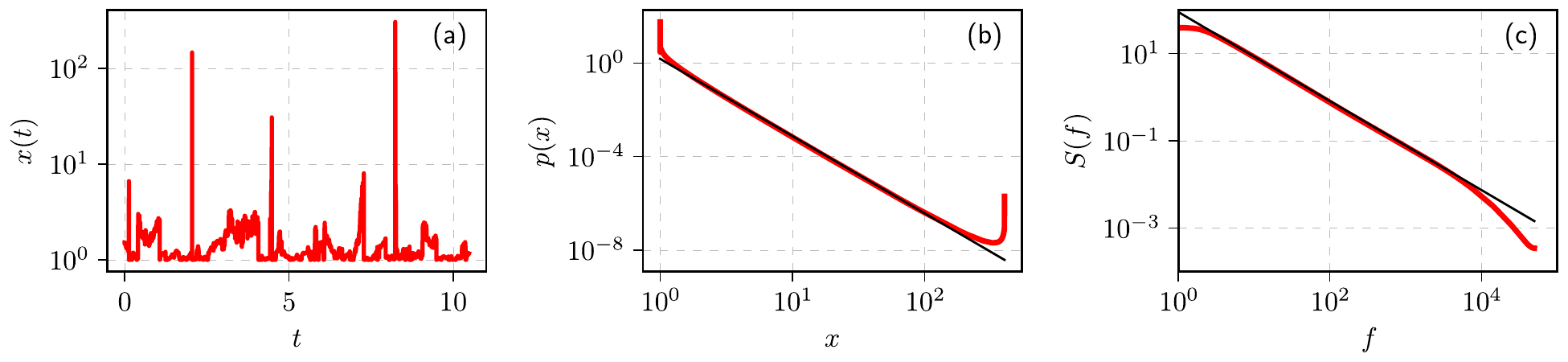}
\caption{Statistical properties of the time series obtained by solving
SDE with L\'evy $\alpha$-stable noise, Eq.~\eqref{eq:sde-levy}: (a)
sample fragment of the time series, (b) PDF and (c) PSD of time series.
Red curves correspond to numerical results, while black curves are
power--law best fits with exponents (b) $\lambda \approx 3.3$, (c) $ \beta
\approx 1$.\label{fig:levy-sde}}
\end{figure}

In Fig.~\ref{fig:levy-sde} we show a sample series of the solutions of
SDE~\eqref{eq:sde-levy} and the statistical properties of the series when
the noise is L\'evy $\alpha$-stable noise with $\alpha=1$. The other
SDE~\eqref{eq:sde-levy} parameters were picked so $1/f$ spectrum would be
reproduced. As can be seen in the subfigure (a) ongoing diffusion is
disrupted by huge jumps, which are characteristic to L\'evy flights.

If we consider modeling only sub--diffusive processes, then we can study
another generalization of SDE~\eqref{eq:sde-class}, originally proposed
in \cite{Kazakevicius2015PhysA}. If we start with a Markovian process
described by the It\^o SDE
\begin{equation}
\rmd x(\tau)= f\left[x(\tau)\right] \rmd\tau+g\left[x(\tau)\right] \rmd
W(\tau).\label{eq:sde-sub-levy}
\end{equation}
The drift and diffusion functions of the above SDE are given by
\begin{equation}
f(x)=\sigma^2 \left(\eta - \frac{\lambda }{2}\right) x^{2\eta-1},
\qquad g(x)=\sigma x^{\eta}.
\end{equation}
We interpret the time $\tau$ as an internal (operational) time. For the
trapping processes that have a distribution of the
trapping times with power--law tails, the physical time $t=T(\tau)$
is given by the strictly increasing $\alpha_{+}$--stable L\'evy motion
defined by the Laplace transform
\begin{equation}
\left\langle e^{-kT(\tau)} \right\rangle=e^{-\tau k^{\alpha_{+}}}.
\end{equation}
Here the parameter $\alpha_{+}$ takes the values from the interval $0<\alpha_{+}<1$.
Thus the physical time $t$ obeys the SDE
\begin{equation}
\rmd t(\tau)=\rmd L^{\alpha_{+}}(\tau),
\end{equation}
where $dL^{\alpha_{+}}(\tau)$ stands for the increments of the strictly
increasing $\alpha_{+}$--stable L\'evy motion $L^{\alpha_{+}}(\tau)$. For
such physical time $t$ the operational time $\tau$ is related to
the physical time $t$ via the inverse $\alpha_{+}$--stable subordinator
\begin{equation}
S(t)=\inf\{\tau:T(\tau)>t\}.
\end{equation}
Such subordination leads to power spectral density
\begin{equation}
S(f)\sim\begin{cases}
\frac{1}{\omega^{\beta}}, & 1-\alpha_{+}<\beta<1+\alpha_{+},\\
\frac{1}{\omega^{1+\alpha_{+}}}, & \beta>1+\alpha_{+}.
\end{cases}
,\qquad\beta=1+\frac{\alpha_{+}(\lambda-3)}{(\eta-1)}
\end{equation}

Proposed SDEs~\eqref{eq:sde-class}, \eqref{eq:sde-levy}, and \eqref{eq:sde-sub-levy}
have served as a basis to study heterogeneous diffusion in non--homogeneous
medium \cite{Kazakevicius2015JStatMech,Kazakevicius2015PhysA,Kazakevicius2016PRE}
and time subordinated processes \cite{Ruseckas2016JStat,Ruseckas2016JStatMech}
as well as the effects of non--linear variable transformations
\cite{Kaulakys2015MPLB,Kazakevicius2021PRE}.

In paper \cite{Ruseckas2016JStat} we investigated the distinction between
the internal time of the system and the physical time as a source of
$1/f$ noise.
We have introduced the internal (operational) time into the
earlier point process
\cite{Kaulakys1998PRE,Kaulakys1999PLA,Kaulakys2005PhysRevE} together with
additional equation relating the internal time to the physical time. In
this scenario we can still recover power--law statistical features similar
to the ones obtained by solving Eq.~\eqref{eq:sde-class}.
In the financial markets, the internal time could reflect the fluctuating
human activity, e.g., trading activity, yielding the long--range correlations
in the volatility.
The effective way for the solution of highly non--linear SDEs was proposed \cite{Ruseckas2016JStat} by suitable choice of the internal time and variable
steps of integration.

The effects of non--linear variable transformations
\cite{Kaulakys2015MPLB,Kazakevicius2021PRE} suggest that long--range memory in
certain cases can be just a measurement effect. As far as the non--linear
transformation of the observable $x$ to $y$
\begin{equation}
x=\frac{1}{y^{\delta}},
\end{equation}
with $\delta$ being the transformation exponent, yields SDE for the variable
$y$ of the same form like Eq. \eqref{eq:sde-class} for $x$.

\subsection{Inverse cubic law for long--range correlated processes}

Inverse cubic law is an established stylized fact stating that the
cumulative distributions of various financial market time series
such as the number of trades, the trade volume or the return
\cite{Gopikrishnan1998EPJB,Cont2001RQUF,Mantegna2000Cambridge,Alfi2009EPJB1}.
Thus this law is as important for the modeling as the consideration of
long--range memory and fractal scaling, which are also stylized facts
\cite{Gopikrishnan1998EPJB,Cont2001RQUF,Mantegna2000Cambridge,Alfi2009EPJB1,Mandelbrot-1999}.
We have proposed \cite{Kaulakys2016JStat} the non--linear SDE giving both
the power--law behavior of the PSD and the inverse cubic law of the cumulative distribution.
This was achieved using the idea that when the market evolves from calm to
violent behavior there is a decrease of the delay time of multiplicative
feedback of the system in comparison to the driving noise correlation time.
This results in transition from the It\^o to the Stratonovich sense of
the SDE and yields a long--range memory process.

We start from a simple quadratic SDE
\begin{equation}
\rmd x=x^2 \circ_\alpha \rmd W
\label{eq:strato-sde-simple}
\end{equation}
where $\alpha$ is the interpretation parameter, defining the
$\alpha$-dependent stochastic integral of the SDE~\eqref{eq:strato-sde-simple},
\begin{equation}
\int_0^T f(x(t))\circ_\alpha dW_t\equiv\lim_{N\to\infty}\sum\limits_{n=0}^{N-1}f(x(t_n))\Delta W_{t_n}.
\label{eq:2}
\end{equation}
Here $t_n=\frac{n+\alpha}{N}T$ with $0 \leq \alpha \leq 1$.
Natural choices of the parameter $\alpha$ are: (i) $\alpha=0$, pre--point
(It\^o convention), (ii) $\alpha=1/2$, mid--point (Stratonovich convention)
and (iii) $\alpha=1$, post--point (H\"{a}nggi--Klimontovich, kinetic or
isothermal convention) \cite{Pesce2013}.

The quadratic SDE~\eqref{eq:strato-sde-simple} is the simplest
multiplicative SDE without the drift term symmetric for the positive and
negative deviations of some observable $x$. More generally, the same process
can be described by the delayed SDE \cite{Pesce2013}
\begin{equation}
\rmd x(t) = f\left( x(t)\right) \rmd t + g\left( x(t-\delta) \right)
\zeta_t^\tau \rmd t. \label{eq:sde-strato-delayed}
\end{equation}
Here $f\left( x \right)$ represents arbitrary deterministic drift
of the observable $x$, while $g\left( x \right)$ effectively controls the
diffusion as $\zeta_t^\tau$ is the noise term, which is assumed to have
correlation time $\tau$. Note that the diffusion function depends on the
delayed value of the observable $x$ (by time interval $\delta$).

It may be shown \cite{Pesce2013} that in the limit $\delta \rightarrow 0$
and $\tau \rightarrow 0$ (under the condition $\delta /\tau =const$)
SDE~\eqref{eq:sde-strato-delayed} can be transformed into
\begin{equation}
\rmd x = f\left( x(t) \right) \rmd t + g\left( x(t) \right) \circ_\alpha \rmd W
\end{equation}
with the interpretation parameter being determined by
\begin{equation}
\alpha \left( \frac \delta \tau \right) \simeq \frac 1{2\left( 1+\delta /\tau \right)}.
\end{equation}

Under the perturbation by the white noise, in a case of $\tau \ll \delta$,
even for short delay in feedback $\delta$ we achieve the It\^o outcome,
because there is no correlation between the sign of the noise $\zeta_t$ and
the time--derivative of the feedback $g\left( x\right)$.
On the contrary, under the perturbation by the correlated noise, $\tau \gg
\delta$, a correlation emerges between the sign of $\zeta_t$ and the
time--derivative of $g\left(x\right)$. In this case the correlation yields
the Stratonovich outcome \cite{Pesce2013}.

In general the value of $\alpha$ may depend on the coordinate $x$ and/or
other system' parameters. SDE~\eqref{eq:strato-sde-simple} with
$\alpha \neq 0$ may be transformed into SDE in It\^o sense
\begin{equation}
\rmd x=2\alpha x^3 \rmd t+x^2 \rmd W. \label{eq:sde-variable-interpretation}
\end{equation}
This SDE is a particular case of the general It\^o equation
\eqref{eq:sde-class} yielding the power--law steady--state PDF and
the power--law PSD \eqref{eq:sde-class-stats}. These SDEs become
identical for $\eta = 2$ and $\lambda = 4 \left( 1 - \alpha \right)$.

Let us note that $1/f^{\beta}$ noise emerges due to the large
fluctuations in the time series, while the finite time studies reveal the
commonly observed magnitudes of the observable. The common fluctuations can
be modeled by the familiar in the financial applications It\^o SDEs.
On the other hand, the large rapid fluctuations of the violent market arise
due to the strong correlated influences, the processes of such a market are
fast, all durations become short in comparison to the herding correlation
time, and, consequently, the market should be modeled by the Stratonovich
version of SDE.

For the modeling of such dynamics we generalize equations
\eqref{eq:strato-sde-simple} and \eqref{eq:sde-variable-interpretation} with
$x$-dependent parameter $\alpha(x)$. Let
\begin{equation}
\rmd x = 2\alpha(x) x^3 \rmd t + x^2 \rmd W, \label{eq:sde-variable-alpha}
\end{equation}
with, e.g.,
\begin{equation}
\alpha(x)=\frac{1}{2}\left[1-\exp\left\{-\left(\frac{x}{x_c}\right)^2\right\}\right],
\label{eq:sde-variable-alpha-def}
\end{equation}
where $x_c$ is the It\^o to Stratonovich interpretations crossover
parameter. Equations \eqref{eq:sde-variable-alpha} and \eqref{eq:sde-variable-alpha-def} represent transition from
It\^o to Stratonovich convention with increasing the variable $x$ and
decrease of the delay time of multiplicative feedback for larger $x$,
according to Wong--Zakai theorem \cite{Pesce2013}.
Detailed numerical analysis of the model represented by equations \eqref{eq:sde-variable-alpha} and \eqref{eq:sde-variable-alpha-def} is presented in paper \cite{Kaulakys2016JStat}.

\subsection{$1/f^\beta$ noise with distributions other than power--law}
\label{sec:1f-other-dist}

Solutions of the SDE~\eqref{eq:sde-class} will always have power--law
statistical properties of the \eqref{eq:sde-class-stats} form. However,
often noise with $1/f^\beta$ PSD is distributed according to PDF, which is
not power--law, but Gaussian or some other distribution. Here we review two
different approaches, which allow for other distributions to be observed in
time series with $1/f^\beta$ spectrum: superstatistical and coupled SDE
approaches.

In \cite{Kaulakys2009BrazJ} it was suggested that the Poissonian--like process
with the slowly changing average inter--event time may be represented as the
superstatistical process one exhibiting $1/f$ noise. It was assumed that the
inter--event time $\tau_k$, obtained by solving Eq.~\eqref{eq:pp-tau}, represents not the actual (observed) inter--event
time, but its average (reciprocal of the event rate). In this setup
the actual inter--event time $\hat{\tau}_k$ would given by the conditional
probability
\begin{equation}
\varphi\left(\hat{\tau}_k \middle|
\tau_k\right)=\frac{1}{\tau_k}e^{-\hat{\tau}_k/\tau_k},
\label{eq:superstats-observed-time-cond}
\end{equation}
like for the non--homogeneous Poisson process. This additional randomization
has no influence on the lower frequencies of the PSD and the intensity of
the signal.

The PDF of the observed inter--event time $\hat{\tau}_k$ may be derived from
the superstatistical model,
\begin{equation}
p\left( \hat{\tau}_k
\right)=\int_0^{\infty}\varphi\left(\hat{\tau}_k\middle|\tau_k\right)p_k(\tau_k) \rmd\tau_k.
\label{eq:superstats-observed-time-pdf}
\end{equation}
Equations~\eqref{eq:superstats-observed-time-cond} and
\eqref{eq:superstats-observed-time-pdf} generate the $q$-exponential
distribution used in the nonextensive statistical mechanics and many real
systems \cite{Tsallis2017Entropy}. Detailed analytical derivations and the
numerical verification was presented in \cite{Kaulakys2009BrazJ}.

In the paper \cite{Ruseckas2012ACS}, a similar superstatistical
approach was taken in respect to the intensity of the signal $x$, obtained by
solving SDE~\eqref{eq:sde-class}. The observed series $\hat{x}$ is assumed
to be generated from $x$ series by applying exogenous noise, which is
described by an arbitrary conditional distribution $\varphi(\hat{x}|x)$.
In such approach the steady--state distribution of $\hat{x}$ is given by
\begin{equation}
p(\hat{x})=\int_0^{\infty}\varphi(\hat{x}|x)p(x) \rmd x.
\label{eq:super}
\end{equation}
Analytical and numerical analysis of inter--trade duration, the trading
activity, and the return using the superstatistical method with the
exponential and normal distributions of the local signal, driven by the
stochastic process, was discussed in detail in \cite{Ruseckas2012ACS}.

Later we have shown that superstatistical approach is not the
only approach, which allows us to change the observed signal PDF. Coupled
SDE approach, proposed in \cite{Ruseckas2016JStatMech}, allows for more
flexibility and easier interpretation of how the statistical properties
become independent of each other. The general form of the set of coupled
SDEs was derived from the scaling properties needed for the
realization of $1/f^\beta$ noise \cite{Ruseckas2016JStatMech}
\begin{align}
\rmd x & = f(x) y^{2\eta} \rmd t + g(x) y^{\eta} \rmd W_{1},
\label{eq:coupled-sde-x} \\
\rmd y & = \sigma^{2}\left(\eta+1-\frac{\lambda}{2}\right)y^{2\eta+1} \rmd t
+\sigma y_{t}^{\eta+1} \rmd W_{2}.
\label{eq:coupled-sde-y}
\end{align}
Here $f(x)$ and $g(x)$ are arbitrary drift and diffusion
functions, which determine the stationary PDF of $x$, $W_{1}$ and $W_{2}$
are uncorrelated standard Wiener processes.
The first equation describes the changes in the intensity of the signal, while
the second equation represents fluctuations in the rate of change. These
coupled SDEs allow for $1/f^\beta$ spectrum to be reproduced together with
arbitrary steady--state PDF of the observed value $x$. It was shown that
the power--law slope of the PSD, $\beta$, of the time series of $x$ generated by
solving SDEs~\eqref{eq:coupled-sde-x} and \eqref{eq:coupled-sde-y} depends
on the parameters $\eta$ and $\lambda$ as follows
\begin{equation}
\beta=1+\frac{\lambda-1}{2\eta}.
\end{equation}

\begin{figure}
\centering
\includegraphics[width=0.9\textwidth]{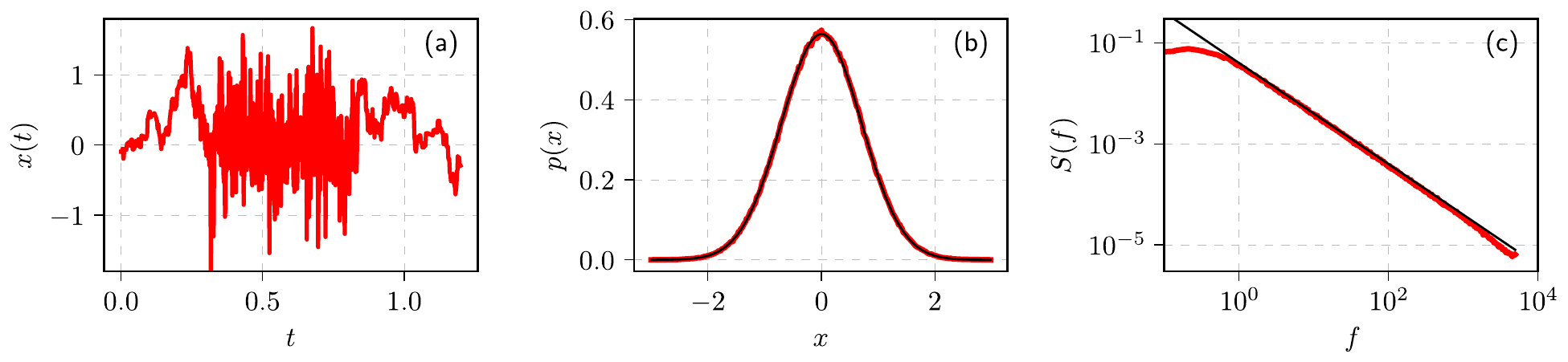}
\caption{Statistical properties of the time series obtained by solving
coupled SDEs \eqref{eq:coupled-sde-x} and \eqref{eq:coupled-sde-y}: (a)
sample fragment of $x(t)$ time series, (b) PDF of the externally observed
values $x$ and (c) PSD of $x(t)$. Red curves correspond to numerical
results, while black curves are theoretical fits: (b) standard Gaussian PDF,
(c) $S(f) \sim 1/f^\beta$.\label{fig:coupled-sde}}
\end{figure}

In Fig.~\ref{fig:coupled-sde} we have shown that one can obtain Gaussian
distribution of $x$ (subfigure (b)) together with $1/f$ spectrum (subfigure
(c)). In subfigure (a) one can visually see the impact of the variations in
$y$, which influences the rate of change of $x$.

\subsection{Reproducing statistical properties of the financial markets}

While qualitatively, the trading activity and the absolute returns
have power--law distributions and exhibit long--range memory property
\cite{Cont2001RQUF,Alfi2009EPJB1}, corresponding empirical statistical
properties have a finer structure. In order to reproduce
the empirical statistical properties in detail some modifications
to the SDE are needed.

\cite{Plerou2000PRE} has determined that Hurst exponents of the trading
activity time series of $1000$ US stocks are remarkably close: $H\approx0.85$.
This implies that PSD of the trading activity should have a power--law
slope $\beta=2H-1\approx0.7$. \cite{Plerou2000PRE} has also investigated
that slope of the PDFs of the trading activity also has a power--law
tail with exponent $\lambda\approx4.4$. It would be impossible to
reproduce such values by using SDE~\eqref{eq:sde-class}, because
Eq.~\eqref{eq:sde-class-stats} implies that if $\lambda>3$, then
$\beta>1$. In our analysis of $26$ US stocks \cite{Gontis2008PhysA}
we have confirmed the slope of the PDF, but we have observed a more
complicated PSD, with two slopes instead of one ($\beta<1$ for both
slopes).

Both of these issues are resolved by a modified SDE for trade intensity,
$n$ \cite{Gontis2007PhysA}:
\begin{equation}
\rmd n=\sigma^{2}\left[\eta-\frac{\lambda}{2}+\left(\frac{n_{0}}{n}\right)^{2}\right]\frac{n^{2\eta-1}}{\left(n\epsilon+1\right)^{2}}\rmd t+\sigma\frac{n^{\eta}}{n\epsilon+1}\rmd W.\label{eq:sde-trade}
\end{equation}
The problem of the two PSD slopes is resolved, because this SDE has
two different effective $\eta$ values. For $n\gg\epsilon^{-1}$ the
effective $\eta$ is equal to the specified parameter value (in the
numerical simulations we have used $\eta=5/2$, thus $\hat{\eta}_{1}=5/2$).
For $n\ll\epsilon^{-1}$ the effective $\eta$ is one smaller than
the specified parameter value $\hat{\eta}_{2}=\eta-1=3/2$). Slope
of the PDF increases from the value predicted in Eq.~\eqref{eq:sde-class-stats}
due to integration, as trading activity is defined as number of trades
per time window $w$, or in the current paramerization an integral
of trade intensity: $N\left[t\right]=\int_{t}^{t+w}n\left(u\right)\rmd u$.

Parameter $n_{0}$ and the related term in the drift function ensure
that $n$ would not get very small as the term causes the potential
to rapidly grow for $n<n_{0}$. This helps us avoid negative trade
intensities, which are impossible by definition, as well as ensure
some level of minimal trading activity, which in our experience may
differ for different stocks and different markets \cite{Gontis2008PhysA,Gontis2011JDySES}.

\begin{figure}
\centering
\includegraphics[width=0.7\textwidth]{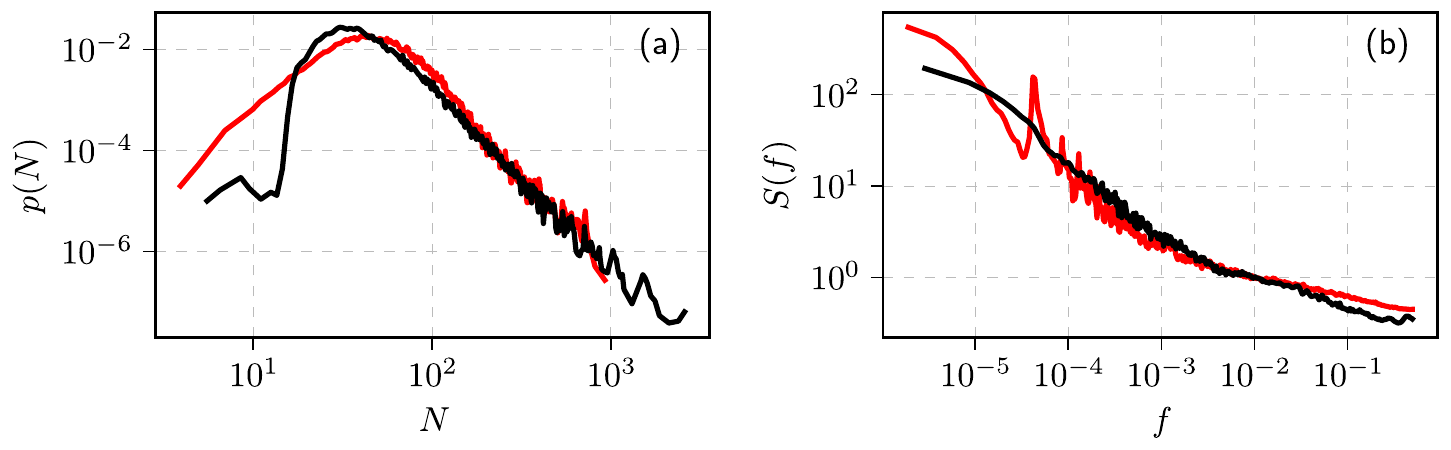}
\caption{Trading activity PDF (a) and PSD (b) for MMM stock traded on NYSE (red curve)
and the numerical solutions of SDE~\eqref{eq:sde-trade}. Model parameters
values: $\eta=2.5$, $\lambda=4.3$, $\sigma^{2}=0.045$, $\epsilon=0.36$,
$n_{0}=0.14$. Empirical and numerical PDF was obtained by considering
trades in the $300\,\mathrm{s}$ time window.\label{fig:sde-trade}}
\end{figure}

In Fig.~\ref{fig:sde-trade} we have shown that the stochastic model
can match statistical properties of MMM stock traded on NYSE.
While the matches are not perfect, but some of the noticeable differences
can be explained by the fact that the stochastic model does not take
into account intraday seasonalities.

Reproducing statistics of absolute return requires another modification
of the SDE \cite{Gontis2010PhysA}. Our empirical analysis, confirmed
by the other authors \cite{Tsallis2017Entropy}, indicated that $q$--Gaussian
distribution \cite{Ruseckas2011PhysRevE,Ruseckas2012ACS} seems to be a good fit for the empirical absolute return,
defined as the log--price difference, distribution. This is achieved
by:
\begin{equation}
\rmd x=\sigma^{2}\left[\eta-\frac{\lambda}{2}-\left(\frac{x}{x_{\mathrm{max}}}\right)^{2}\right]\frac{\left(1+x^{2}\right)^{\eta-1}}{\left(1+\epsilon\sqrt{1+x^{2}}\right)^{2}}x\rmd t+\sigma\frac{\left(1+x^{2}\right)^{\frac{\eta}{2}}}{1+\epsilon\sqrt{1+x^{2}}}\rmd W.\label{eq:sde-return}
\end{equation}
To reproduce the full complexity of the empirical data another ingredient
is needed: external noise, which can be understood as an effect of
news flow or the distortions caused by the discrete order flow:
\begin{equation}
r_{t}=\xi\left\{ r_{0}=1+\frac{2}{w}\left|\int_{t-w}^{t}x\left(u\right)\rmd u\right|,q=1+2/\lambda_{2}\right\} .\label{eq:ret-noise}
\end{equation}
This relation was inspired by the superstatistical approach
(discussed in Section~\ref{sec:1f-other-dist}) and determined by trying to
fit the empirical data as
best we can. We have empirical determined that the best fit is obtained
when $\xi$ is a process that generates uncorrelated random variates
from a $q$--Gaussian distribution with $q\approx1.4$ ($\lambda_{2}\approx5$)
and $r_{0}$ being one minute ($w\approx60\,\mathrm{s}$) moving average
filter of the solutions of SDE~\eqref{eq:sde-return}. Using this
model, we were able to reproduce empirical statistical properties of
stock from New York (abbr. NYSE) and Vilnius (abbr. VSE) stock exchanges
\cite{Gontis2010PhysA,Gontis2011JDySES}.

\begin{figure}
\centering
\includegraphics[width=0.7\textwidth]{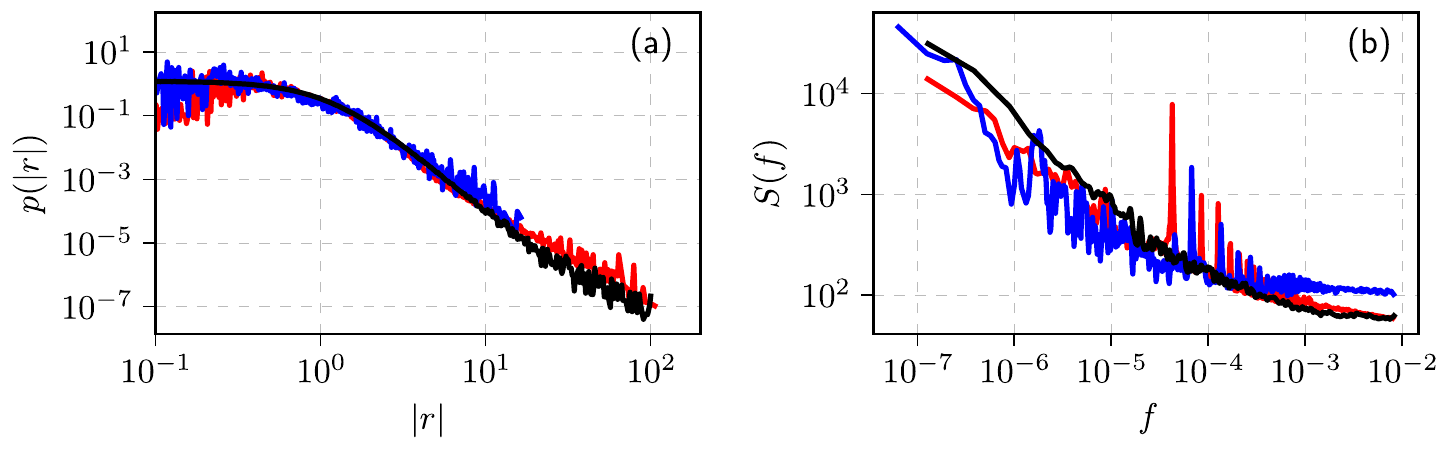}
\caption{Comparison of empirical PDF (a) and PSD (b) of absolute one minute return
as observed in NYSE (red curves) and VSE (blue curves) stocks. Empirical
results are compared against the model, generated by the SDE~\eqref{eq:sde-return}
and exogenous noise Eq.~\eqref{eq:ret-noise}, (black curves). Model
parameter values: $\eta=2.5$, $\lambda=3.6$, $\epsilon=0.017$,
$x_{\mathrm{max}}=10^{3}$, $\lambda_{2}=5$.\label{fig:sde-return}}
\end{figure}

In Fig.~\ref{fig:sde-return} we have demonstrated that the stochastic model reasonably well reproduces empirical data from NYSE and VSE. Some of the noticeable differences can be observed because
we do not take into account the intraday seasonality, and we do not
directly take into account that VSE had relatively low liquidity (many
one minute time intervals have zero returns). Differing liquidity
is a likely explanation for the differences seen between NYSE and
VSE, too.

\subsection{Variable step method for solving non--linear stochastic differential
equations}

Note that SDEs \eqref{eq:sde-class}, \eqref{eq:sde-trade} and \eqref{eq:sde-return}
are not Lipshitz continuous \cite{Kloeden1999Springer}; thus, they
have to be solved by imposing boundary conditions, which would prevent
the explosion of the solutions. An alternative way to achieve Lipshitz
continuity is to include additional terms for restricting diffusion,
which would have no detrimental effects on the PSD and PDF of the
time series. Such is the role of the $n_{0}$ term in SDE~\eqref{eq:sde-trade}
and $x_{\mathrm{max}}$ term in SDE~\eqref{eq:sde-return}.

Lacking Lipshitiz continuity causes another complication in solving
the SDEs: the standard Euler--Maruyama or Milsten methods \cite{Kloeden1999Springer}
do not yield good results with reasonable step sizes. This complication
is resolved by using variable step size. The core idea is to use
a larger step size whenever the anticipated changes would be small
and use the smaller step size whenever significant changes are coming. The mathematical
form of the variable step size is often unique to the SDE being
solved, but a good rule of thumb would be to linearize the drift
and the diffusion functions. See \cite{Kaulakys2004PhysRevE,Kaulakys2006PhysA}
for more details.

For example, SDE~\eqref{eq:sde-class} in our works is solved by the
following set of difference equations:
\begin{align}
x_{i+1} & =x_{i}+\kappa^{2}\left(\eta-\frac{\lambda}{2}\right)x_{i}+\kappa x_{i}\varepsilon_{i},\\
t_{i+1} & =t_{i}+\kappa^{2}x^{2-2\eta}.
\end{align}
In the above $\kappa$ is a small number that acts as an error tolerance
parameter. The smaller it gets the better $x_{i}$ reproduces desired statistical properties given by Eq.~\eqref{eq:sde-class-stats},
but at the expense of numerical computation time.

Similarly this variable step method can be also applied to SDEs with
$\alpha$--stable L\'evy noise. For example, we can solve SDE~\eqref{eq:sde-levy}
numerically by using the following set of difference equations
\begin{align}
x_{k+1} & =x_{k}+\kappa^{\alpha}\gamma
x_{k}+\frac{\kappa}{\sigma}x_{k}\xi_{k}^{\alpha}, \\
t_{k+1} & = t_{k} + \frac{\kappa^{\alpha}}{\sigma^{\alpha}}x_{k}^{-\alpha(\eta-1)},
\end{align}
Here $\xi_{k}^{\alpha}$ is a random variable having
$\alpha$--stable L\'evy distribution. This set of difference equations should
be solved only with the reflective
boundaries at $x=x_{\mathrm{min}}$ and $x=x_{\mathrm{max}}$ using
the projection method \cite{Pettersson1995}. In nutshell,
if the variable $x_{k+1}$ acquires the value
outside of the interval $[x_{\mathrm{min}},x_{\mathrm{max}}]$ then
the value of the nearest reflective boundary is assigned to $x_{k+1}$.
Iterative equations for SDEs \eqref{eq:sde-trade} and \eqref{eq:sde-return}
are a bit more complicated \cite{Gontis2008PhysA,Gontis2010PhysA},
but they still remain qualitatively the same.

Note that introduction of the variable time step into the
numerical solution of
an SDE is equivalent to introducing the subordination scheme directly into
the SDE, when internal time and physical time are related by a non--linear
transformation \cite{Ruseckas2016JStat}.

\section{Agent--based model of the long--range memory in the financial markets\label{sec:abm}}

In the previous section, we have discussed how our group has started
from the physically motivated point process model and arrived at the
general class of SDEs reproducing long--range
memory phenomenon. However, this generality has its drawback: microscopic
mechanisms of the modeled systems are ignored. We then tried to investigate
some existing financial ABMs for the possibility to derive SDE of
a similar form to SDE~\eqref{eq:sde-class}. We have failed to do
so with some prominent yet complicated ABMs like the ones proposed
in \cite{Lux1999Nature,Challet2000PhysRevLett} (for more prominent
ABMs of the time, which include some other candidates we have tried,
see \cite{Cristelli2012Fermi}). However, we have found success with Kirman's
herding model, initially proposed in \cite{Kirman1993QJE} and later
analyzed in financial market context by \cite{Alfarano2005CompEco,Alfarano2008Dyncon}.

\subsection{Kirman's herding model}

Kirman's herding model can be defined via two one--step transition
probabilities in a system with two possible states:
\begin{align}
p\left(X\rightarrow X+1\right) & =\left(N-X\right)\left[\sigma_{1}+hX\right]\Delta t,\\
p\left(X\rightarrow X-1\right) & =X\left[\sigma_{2}+h\left(N-X\right)\right]\Delta t.
\end{align}
In the above $X$ is the number of agents in state $1$ and $N$ is
the total number of agents within the system. Total number of agents
is conserved, so the number of agents in the state $2$ is trivially
given by $N-X$. Here $\Delta t$ is a short time window during which
only one transition should be likely. Transitions may occur either
due to independent behavior (governed by parameters $\sigma_{i}$),
or due to recruitment (governed by parameter $h$). Using birth--death
process formalism \cite{VanKampen2007NorthHolland} it is easy to
find SDE corresponding to Kirman's herding model with $x=X/N$:
\begin{equation}
\rmd x=\left[\left(1-x\right)\sigma_{1}-x\sigma_{2}\right]\rmd t+\sqrt{2hx\left(1-x\right)}\rmd W.\label{eq:sde-kirman-x}
\end{equation}

\subsection{Kirman's herding model for the financial markets}

Evidently SDE~\eqref{eq:sde-kirman-x} is not of the same form as
SDE~\eqref{eq:sde-class}, but we have not yet discussed the meaning
of the states $1$ and $2$. In many financial ABMs of the time it
was a common choice to assume that agents represent chartist and fundamentalist
traders \cite{Cristelli2012Fermi}. Assuming that chartist traders
trade based on the wide variety of technical trading tools, which
often produce conflicting predictions, their excess demand (difference
between the supply and demand generated by the group as a whole) is
given by:
\begin{equation}
D_{c}=r_{0}X_{c}\left(t\right)\xi\left(t\right),
\end{equation}
where $X_{c}\left(t\right)$ is the number of chartist traders and
$\xi\left(t\right)$ is their average mood (describing average sentiment
to buy or sell). The relative impact of the chartists' traders in comparison
to fundamentalist traders is given by $r_{0}$. Fundamentalist traders
on the other hand, they are often assumed to trade based on the quantity
known as a fundamental price, $P_{f}$, with the expectation that the price,
$P\left(t\right)$, in the long run, will converge towards the fundamental
price. Under this assumption, their excess demand is given by:
\begin{equation}
D_{f}=X_{f}\left(t\right)\ln\frac{P_{f}}{P\left(t\right)}.
\end{equation}
Using the excess demand functions of the both groups, we can use Walras
law \cite{Walras2013} to obtain the expression for the price\cite{Alfarano2005CompEco,Kononovicius2012PhysA}:
\begin{equation}
P\left(t\right)=P_{f}\exp\left[r_{0}\frac{X_{c}\left(t\right)}{X_{f}\left(t\right)}\xi\left(t\right)\right].\label{eq:abm-eq-price}
\end{equation}
Log--return of the price is evidently given by:
\begin{equation}
r_{w}\left(t\right)=\ln P\left(t\right)-\ln P\left(t-w\right)=r_{0}\frac{x_{c}\left(t\right)}{x_{f}\left(t\right)}\zeta_{w}\left(t\right).
\end{equation}
In the above $\zeta_{w}\left(t\right)$ is the mood change function
over time window $w$. As the mood changes on a very short time scale
and we are interested in the long--term dynamics we can simply assume
that $\zeta_{w}\left(t\right)$ is some kind of uncorrelated noise
and consider only a more slowly varying ratio between fractions of
chartists and fundamentalists. As the total number of agents is fixed
we can define long--term component of return, modulating return,
as:
\begin{equation}
y\left(t\right)=\frac{x\left(t\right)}{1-x\left(t\right)}.
\end{equation}
SDE for the modulating return is given by:
\begin{equation}
\rmd y=\left[\sigma_{1}+\left(2-\sigma_{2}\right)y\right]\left(1+y\right)\rmd t+\sqrt{2hy}\left(1+y\right)\rmd W,
\end{equation}
which is roughly similar to the SDE~\ref{eq:sde-class} with $\eta=3/2$
and $\lambda=\frac{\sigma_{2}}{h}+1$.

This SDE can be generalized by introducing variable event rate $\tau\left(y\right)=y^{-\alpha}$.
This addition can be explained by the fact that it is well known that
returns and trading volume correlate and the best correlation is acchieved
between squared returns and volume \cite{Gabaix2003Nature,Farmer2004QF,Gabaix2006QJE,Rak2013APP},
hence suggesting that $\alpha=2$ is a likely candidate. With this
extension and when considering only the highest powers of $y$ (as
the large $y$ tend to influence the PSD), we obtain \cite{Kononovicius2012PhysA}:
\begin{equation}
\rmd y=h\left(2-\sigma_{2}\right)y^{2+\alpha}\rmd t+\sqrt{2hy^{3+\alpha}}\rmd W.
\end{equation}
Now this SDE is completely equivalent to the SDE~\ref{eq:sde-class}
with $\eta=\frac{3+\alpha}{2}$ and $\lambda=\frac{\sigma_{2}}{h}+\alpha+1$.
Consequently PSD of $y$ will have a frequency range in which:
\begin{equation}
S_{y}\left(f\right)\sim1/f^{\beta},\quad\beta=1+\frac{\frac{\sigma_{2}}{h}+\alpha-2}{1+\alpha}.
\end{equation}

In the later papers, we have modified this herding ABM until it was
able to reproduce absolute return PDF and PSD close to the empirical
absolute return PDFs and PSDs. In \cite{Kononovicius2013EPL} we have
shown that considering mood dynamics can help in reproducing fractured
PSD. In \cite{Gontis2014PlosOne} we have reliably introduced the
exogenous noise, much similar to what was done with SDE driven model
in \cite{Gontis2010PhysA}, into this ABM, thus producing a consentaneous
model. In \cite{Kononovicius2014PhysA,Kononovicius2015EPJB} we have explored the opportunities
to control the fluctuations in the artificial financial markets driven
by the herding ABM, showing that random trading, control strategy
suggested in \cite{Biondo2013PhysRevE}, may also destabilize the
market. In \cite{Kononovicius2019OB} we have removed the assumption
about the exogenous noise and replaced it with order book dynamics,
thus presenting another possible explanation for fracture in the PSD:
it also arises due to market price lagging behind the changes in the
equilibrium price, Eq.~\ref{eq:abm-eq-price}. Notably, the order
book version of the model was able to reproduce both trading activity
and absolute return statistical properties at the same time.

\begin{figure}
\centering
\includegraphics[width=0.7\textwidth]{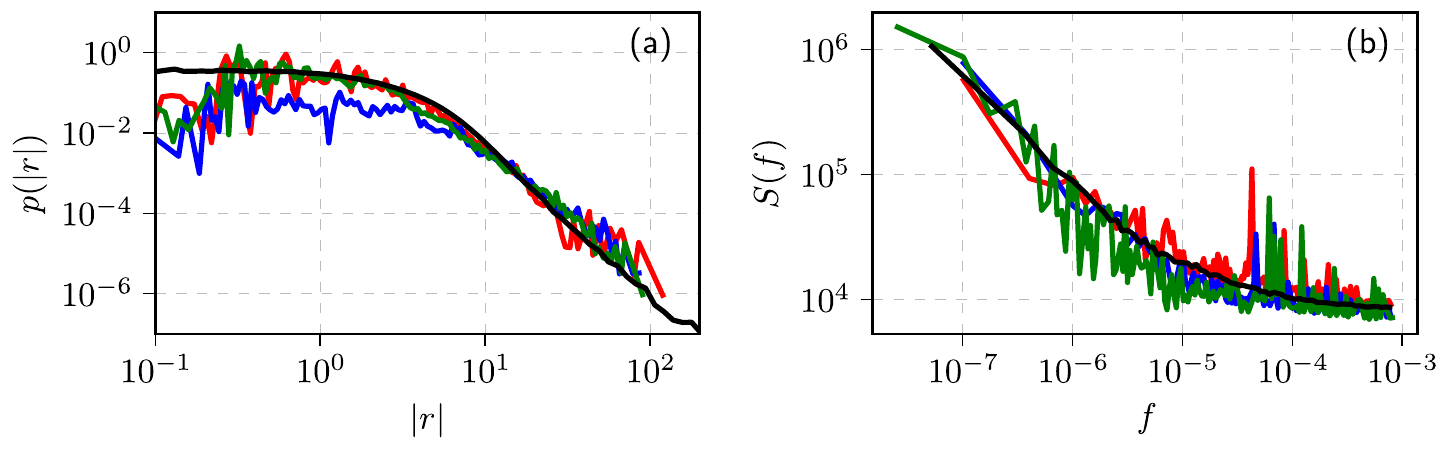}
\caption{Comparison of empirical PDF (a) and PSD (b) of absolute ten minute return
as observed in NYSE (red curves), VSE (blue curves) and WSE (green
curves) stocks. Empirical results are compared against the consentaneous
model, defined in \cite{Gontis2014PlosOne}. Model parameter values
are the same as in Fig.~2 of \cite{Gontis2014PlosOne}.\label{fig:abm-return}}
\end{figure}

In Fig.~\ref{fig:abm-return} we have reproduced one of the figures
from \cite{Gontis2014PlosOne} to show how well the ABM can reproduce the
empirical data from New York, Vilnius, and
Warsaw (abbr. WSE) stock exchanges. Here
we have shown that the model was able to reproduce $10$ minute absolute
return PDFs and PSDs from the different stock exchanges, but in the
original article, more intraday time scales are covered, and seasonality
was also taken into account.

\subsection{Kirman's herding model, voter model and the opinion dynamics context}

Attentive reader with a background in opinion dynamics will likely
notice that Kirman's model is remarkably similar to the well--known
voter model \cite{Castellano2009RevModPhys,Dong2018IntFus,Noorazar2020EPJP}.
They are identical, which has prompted us to question whether
the voter model is truly a model for voters, which Fernandez--Garcia
et al. in \cite{FernandezGarcia2014PRL} also raised. This has lead us to explore
and model statistical properties of spatially heterogeneous electoral
data \cite{Kononovicius2017Complexity}. As we have noticed segregation
effects in the electoral data, we have continued our investigation
by considering the migratory nature of census and electoral data \cite{Kononovicius2019CompJStat}.
Similar approaches were taken by others as well. Sano and Mori \cite{Sano2016}
have looked into spatiotemporal Japanese election data in their
model, assuming a noticeable fraction of stubborn voters
who do not allow for the party's popularity to drop below a certain
threshold. Braha et al. \cite{Braha2017PlosOne} have considered spatiotemporal
US election data and have also emphasized the role of opinion leaders
and spatial variability of external influences. Fenner et al. \cite{Fenner2017,Fenner2017QQ}
have started from a generative model inspired by survival analysis,
but in later works transition to the SDE framework \cite{Fenner2018JPhysComm,Levene2021IJF}.
While Michaud and Szilva \cite{Michaud2018PRE} have fixed issues
with the model originally proposed by Fernandez--Garcia et al. \cite{FernandezGarcia2014PRL},
mainly they have redefined how the noise term is handled so that
the model would be more mathematically well--posed. Marmani et al.
\cite{Marmani2020Entropy} have provided a similar empirical analysis
of Italian electoral data and provided additional perspective from
the point of view of Shannon entropy.

As is common in opinion dynamics \cite{Castellano2009RevModPhys,Dong2018IntFus,Noorazar2020EPJP}
we have also explored the influence of network topologies on the statistical
properties of Kirman's herding model. Namely, we have demonstrated
\cite{Kononovicius2014EPJB} a continuous transition from extensive
case, characterized by localized interactions, Gaussian distributions
and Boltzmann entropy, to a non--extensive case, characterized by
global interactions, $q$--Gaussian distribution, and Tsallis entropy.
Similar results were demonstrated earlier by Alfarano and Milakovic
\cite{Alfarano2009Dyncon}, who have explored how Kirman's herding
model works on random, Barabasi--Albert and small--world network
topologies. Similar observations were also made in \cite{Carro2016},
but Carro et al. have used so--called annealed approximation which
takes into account network structures better than the usual mean--field approximation.

Recently we have also used the noisy voter model to model parliamentary presence
\cite{Kononovicius2020JStatMech}. A paper by Vieira et al. \cite{Vieira2019PRE}
has inspired us to look into the Lithuanian parliamentary presence
data. Unlike Vieira et al., we have observed not a ballistic diffusion
regime but superdiffusive behavior. However, both of these regimes can
be obtained from the noisy voter model with imperfectly acting agents.
Namely, agents can internally intend to attend the parliamentary session
or skip, but the action itself may be random despite being conditioned
on the intended action. As Vieira et al. have used fractional diffusion
equation as a model, this result implies that it may be possible to
fake long--range memory encoded in fractional diffusion equation by using Markov
models employing non--linear transformations of the voter model
\cite{Kazakevicius2021PRE}.

Classical voter model incorporates only recruitment
mechanism, while other responses to social interaction are also possible.
For example, diamond model \cite{Willis1965} posits that independence and
anti--conformity mechanism may be important to understanding human social
behaviors. Similarly Latane social impact theory \cite{Latane1981} perdicts
importance of supportive interactions. Namely, individuals strengthening
the conviction of their like--mended peers. While this theory was recently
studied in the opinion dynamics context \cite{Bancerowski2019EPJB,
Kowalska2020PLOS}, it hasn't been combined with the voter model. One could
also consider majority--vote models
\cite{Oliveira1992JStatPhys,Vilela2018SciRep,Galesic2019PhysA}
and q--voter models \cite{Castellano2009PRE,Jedrzejewski2019CRP} as
implementing some kind of support by the like--minded agents. In
majority--vote models recruitment is only possible if majority of agents
have opposing opinion (therefore majority becomes harder to convince, but
minority remains as susceptible to change). In most q--voter models a group
of $q$ agents must share an opinion to convince a single agent. We have
implemented supportive interactions by decreasing the transitions rates of
the agents by an amount proportional to the number of like--minded agents.
In some cases these modifications cause the transition rates go to zero,
which freezes the system state. Similar qualitative behavior is observed in
works, which consider non--Markovian mechanisms, such as implicit opinion
freezing or ageing
\cite{Stark2008PRL,Stark2008ACS,Wang2014SciRep,Artime2018PRE}. This serves
as another example that highly non--linear Markovian models can lead to
similar dynamics as the dynamics generated by the non--Markovian models.

\section{Searching for the true long--range memory test\label{sec:arfima}}

We have reviewed our experience of
modeling long--range memory phenomena using Markovian models in the earlier sections. We have shown
numerous examples of non--linearity causing behaviors and dynamics
reminiscent of the models with true long--range memory (such as delayed
feedback, aging, freezing, and fractional dynamics). In this
section, we present our latest endeavor to find a statistical test, which
would distinguish whether the real--life systems possess true or
spurious long--range memory. We have earlier proposed a test based on the
specific first--passage times, which we refer to as the burst and
inter--burst duration analysis (abbr. BDA)
\cite{Gontis2012ACS,Gontis2017Entropy,Kononovicius2019BDJStat,Gontis2020PhysA}.

Investigating empirical PDF of
burst and inter--burst duration compared with the model properties, we have
interpreted the observed long--range memory in the financial markets by
ordinary non--linear SDEs representing multifractal stochastic processes
with non--stationary increments \cite{Gontis2017PhysA,Gontis2018PhysA}. One has to take into account the interplay of endogenous and exogenous fluctuations in the financial markets to build a comprehensive model of this complex system \cite{Gontis2016APPA}.
Non--linear SDEs might be applicable in the modeling of other social
systems, where models of opinion or populations dynamics lead to the
macroscopic description by these  equations
\cite{Gontis2012ACS,Gontis2017Entropy,Kononovicius2019BDJStat,Gontis2020PhysA}.
The description by SDEs is an alternative to the modeling incorporating
fractional dynamics, if power--law statistical properties are observed in the empirical data.

The BDA
employs the dependence of first--passage time PDF on Hurst
exponent $H$ for the fractional Brownian motion \cite{Ding1995fbm,Metzler2014WorldScientific,Gontis2017PhysA,Gontis2018PhysA}.

FBM, FLSM, and ARFIMA
\cite{Burnecki2010PRE,Burnecki2014JStatMech,Burnecki2017ChaosSF} form the
theoretical background of long--range memory and self--similar processes.
These processes, first of all, served for the modeling of systems with
anomalous diffusion and expected fractional dynamics
\cite{Klafter2012WorldScientific}. We can consider fractional models
possessing true long--range memory
as they have correlated increments. Self--similar processes with non--Gaussian
stable increments are essential for the modeling of social systems as well.
In the financial markets
power--law distributions of noise often interplay with autocorrelations
\cite{Lillo2004SNDE,Bouchaud2004QF,Toth2015JEDC}. In \cite{Gontis2020JStat}
we implemented BDA for the
order disbalance time series seeking to confirm or reject the long--range
memory in the order flow. Further, we analyzed the same LOBSTER data of
order flow in the financial markets \cite{Huang2011Lobster} from the
perspective of FLSM and ARFIMA models seeking to identify the impact of
increment distributions and correlations on estimated parameters of
self--similarity \cite{Gontis2021Arxiv}. The revealed peculiarities of
non--Gaussian fractional dynamics in this financial system raise new
questions about whether used sample estimators are reliable. In this
section, we test various long--range memory estimators such as Mean squared displacement, Absolute Value estimator, Higuchi's method, and BDA  on discrete fractional L\`{e}vy stable motion represented by the ARFIMA sample series.

\subsection{Fractional processes with non--Gaussian noise}
FBM serves as a model of the correlated time series with stationary Gaussian
increments and generalizes the classical Brownian motion
\cite{Mandelbrot1968SIAMR}. One can define FBM, $B_H(t)$, of the  index $H$
(Hurst parameter) in the interval $0<H<1$  as the It\^o integration over classical Brownian motion $B$
\begin{equation}
B_H(t)=\int_{-\infty}^{\infty}\left((t-u)_+^d-(-u)_+^d\right) \rmd B(u),
\end{equation}
where $d=H-1/2$, $(x)_+=\max \left( x,0 \right)$. The parameter $H$ in FBM
quantifies fractal behavior, long--range memory, and anomalous diffusion.
This is not the case for the other more general stochastic processes. Thus
in this contribution  the Hurst parameter $H$ is responsible only for the
fractal properties of the trajectories. We will consider fractional L\`{e}vy stable motion as
more general process with non--Gaussian distribution $L_H^{\alpha}(t)$ representing an integrated process of independent and stable stationary increments $\rmd L^{\alpha}(u)$ \cite{Burnecki2010PRE}
\begin{equation}
L_H^{\alpha}(t)=\int_{-\infty}^{\infty}\left((t-u)_+^d-(-u)_+^d\right) \rmd L^{\alpha}(u),
\end{equation}
where parameter $d$ depends on $H$ and parameter of stable distribution
$\alpha$, $d=H-1/\alpha$. The parameter $\alpha$ characterizes special class
of stable, invariant under summation, distributions
\cite{Smarodinsky1994ChapmanHall}, useful in the modeling both super and
sub--diffusion \cite{Klafter2012WorldScientific}. Here we are interested in the symmetric zero mean, stable distribution defined by the stability index in the region $0<\alpha<2$. This new parameter is responsible for the power--law tails of the new PDF $P(x) \sim \vert x \vert^{-1-\alpha}$.

FBM and FLSM exhibit identical self--similar scaling behavior in
statistical sense,
\begin{equation}
B_H(c t) \sim c^H B_H(t) , \qquad L^\alpha_H(c t) \sim c^H L^\alpha_H(t),
\end{equation}
here $x \sim y$ means that $x$ and $y$ have identical distributions.
One can
establish the relation with the fractal dimension of trajectories $D=2-H$
\cite{Magdziarz2013JPhysA}. In analogy to the notions used in fractal
geometry, these types of processes can be considered self--similar.

Mean squared displacement (abbr. MSD) is another important
statistical property of various complex systems. Mathematically it was introduced as an ensemble  average of the possible microscopic trajectories $x(t)$
\cite{Klafter2012WorldScientific}
\begin{equation}
\langle \left( x(t) - x(0) \right)^2 \rangle \sim t^{\lambda}, \quad  \lambda=2d+1.
\label{eq:MSD}
\end{equation}
Note that Eq. \eqref{eq:MSD} is valid for the FBM, while the ensemble average
of FLSM diverges \cite{Burnecki2010PRE}. For the FBM $d=H-1/2$, while for the
FLSM $\lambda$ is not defined. When $d<0$, one observes dynamics as
sub--diffusion and for $d>0$ as super--diffusion.

In experimental or empirical data analysis one usually deals with discrete--time sample data series $\{X_i\}$. It is challenging to decide which model to apply in the description of empirical data when diffusion is anomalous $d \neq 0$, as observed dynamics in the sample data can originate from the long--range memory or power--law of the noise. We will use the sample MSD defined as
\begin{equation}
M_N(k)=\frac{1}{N-k+1} \sum_{i=o}^{N-k} (X_{i+k}-X_k)^2 . \label{eq:sample-MSD}
\end{equation}

Let us also introduce increment process $\{Y_i=X_i-X_{i-1}\}$
which is extracted from the sample data series. In the case of the FBM
increment process is called fractional Gaussian noise (abbr. FGN), and in
the case of FLSM is called fractional L\`{e}vy stable noise (abbr. FLSN).
Authors in \cite{Burnecki2010PRE} provide an evidence of FLSM non--ergodicity and that $M_N(k) \sim k^{\lambda}$, where $\lambda=2d+1$, for large $N$, $k$, and $N/k$. Thus the MSD sample analysis of  time series with FLSM assumption becomes very important providing estimation of the memory parameter $d$.
The long--range memory usually is defined through the divergence of autocovariance $\rho(k)$, $\sum_{k=1}^{\infty} \rho(k) = \infty$, \cite{Taqqu1995Fractals}
\begin{align}
\rho(k)=\frac{1}{N-k+1} \sum_{i=1}^{N-k+1} Y_i Y_{i+k}=2^{-1} \lbrace (k+1)^{2H}-2k^{2H}+\vert k-1\vert^{2H}\rbrace \\
\sim H(2H-1)k^{-\gamma},\quad k \rightarrow \infty. \nonumber
\label{eq:FGN-autocorrel}
\end{align}
For the FGN, the exponent of autocorrelation is defined by the Hurst parameter $\gamma = 2-2H$. We see that FBM is an essential long--range memory process with various statistical properties defined by the Hurst parameter. Thus, researchers use an extensive choice of statistical estimators to determine $H$ and evaluate memory effects even when investigated time series deviate from the Gaussian distribution.

Accepting more general FLSM approach one has to reevaluate previously used
estimators \cite{Gontis2020JStat}, as we now have more independent
parameters. The stability index $0<\alpha<2$ and the memory parameter $d$
both contribute to the observed sample properties. Since in the L\`{e}vy
stable case the second moment is infinite the measure of noise
autocorrelation, e.g., the co--difference \cite{Smarodinsky1994ChapmanHall,Weron2005PRE}, is used instead of covariance
\begin{equation}
\tau(k) = \sim k^{-(\alpha-\alpha H)}.
\end{equation}
Note that the parameter $\gamma = \alpha-\alpha H = \alpha-\alpha d-1$, has a strong dependency on $\alpha$, when for the Gaussian processes, it was considered just as the indicator of long--range memory. Consequently, the previously used sample power spectral density analysis, the rescaled range analysis \cite{Hurst1951,Beran1994Chapman,Montanari1999MCM}, or multifractal detrended fluctuation analysis \cite{Peng1994PRE,Kantelhardt2002PhysA} has to be reevaluated from the perspective of FLSM \cite{Gontis2020JStat,Gontis2021Arxiv}.

Earlier, we have introduced the burst and inter--burst duration analysis
(BDA) as one more method to quantify the long--range memory through the
evaluation of $H$
\cite{Gontis2017PhysA,Gontis2017Entropy,Gontis2018PhysA,Gontis2020JStat}.
For the one dimensional bounded sample time series, any threshold divides
these series into a sequence of burst $T_j^b$ and inter--burst $T_j^i$
duration, $j=1,..N_b$. The notion of burst and inter--burst duration follows
from the threshold first--passage problem initiated at the nearest vicinity
of the threshold. The burst duration is the first--passage time from above
and inter--burst from below the threshold, see
\cite{Gontis2020JStat,Gontis2018PhysA,Gontis2017PhysA,Gontis2017Entropy} for
more details. The empirical (sample) PDF (histogram) of $T_j$ gives us the
information about $H$, as power--law part of this PDF should be $T^{2-H}$ \cite{Ding1995fbm}. We have to revise the method of BDA from the more general perspective of FLSM \cite{Gontis2021Arxiv}, as the question of which properties can be recovered using this method is open and has to be investigated.

The method of Absolute Value estimator (abbr. AVE) works correctly even for the
time series with infinite variance
\cite{Taqqu1995Fractals,Mercik2003APP,Weron2005PRE,Magdziarz2013JPhysA}. The
method is based on mean value $\delta_n$ calculated from sample series $Y_i$
and evaluating its scaling with length of sub--series $n$. Divide the increment series $Y_i$
into blocks of size $n$, so
that $m \cdot n=N$, and average within each block to get aggregated series
$Y_j^{(n)}=\frac{1}{n} \sum_{i=(j-1)n+1}^{jn} Y_i$. Calculate $\delta_n$
\begin{equation}
\delta_n = \frac{1}{m} \sum_{j=1}^m \vert Y_j^{(n)}-\langle Y \rangle\vert,
\label{eq:AVE}
\end{equation}
where $\langle X \rangle$ is the overall series mean. Then the  absolute value scaling parameter $H_{AV}$ can be evaluated from the scaling relation
\begin{equation}
\delta_n \sim n^{H_{AV}-1}.
\label{eq:AVE-scaling}
\end{equation}

One more almost equivalent estimator of scaling properties regarding the FLSM is Higuchi's method \cite{Higuchi1988PhysD,Taqqu1995Fractals}. It relies on finding fractional dimension $D$ of the length of the path. The normalized path length $L_n$ in this method is defined as follows
\begin{equation}
L_n = \frac{N-1}{n^3} \sum_{i=1}^n \frac{1}{m-1} \sum_{j=1}^{m-1}\vert X_{i+jn}-X_{i+(j-1)n}\vert,
\label{eq:Higuchi}
\end{equation}
and $L_n \sim n^{-D}$, where $D=2-H$.

We will investigate four methods: AVE, Higuchi's, MSD, and BDA for the analysis of ARFIMA time series as a test sample of FLSM.

\subsection{Numerical exploration of the accumulated
ARFIMA(0,d,0) time series}

Let us consider discrete process $\{X_i\}$ defined as a
cumulative sum,
\begin{equation}
    X_{i+1} = X_{i} + Y_i , \label{eq:ARFIMA-acc-unbounded}
\end{equation}
of correlated increments $\{Y_i\}$. Let the increments be generated by the
ARFIMA(0,d,0) process \cite{Stoev2004Fractals,Burnecki2017ChaosSF}:
\begin{equation}
Y_i= \sum_{j=0}^{\infty} \frac{\Gamma(j+d)}{\Gamma(d) \Gamma(j+1)} Z_{i-j},
\label{eq:ARFIMA-increments}
\end{equation}
with random $Z_{i-j}$ from the domain of attraction of an $\alpha$--stable
law with $0 < \alpha \leq 2$. One can calculate the sum in Eq.
\eqref{eq:ARFIMA-increments} using the Fast Fourier Transform algorithm. The
approximate relation between FLSM and ARFIMA can be derived using Riemann--sum approximation, see \cite{Stoev2004Fractals} for details.

\begin{figure}
\centering
\includegraphics[width=0.9\textwidth]{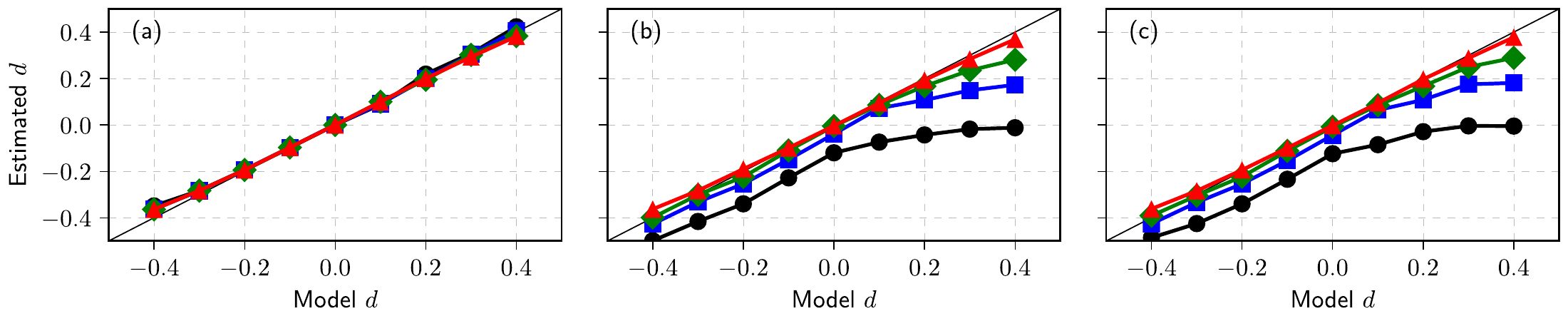}
\caption{Comparison of the MSD (a), AVE (b) and Higuchi (c)
estimator performance when estimating $d$ from the accumulated ARFIMA(0,d,0)
series in the unbounded case, $\{X_i\}$ generated by Eq.~\eqref{eq:ARFIMA-acc-unbounded}.
Different curves correspond to the different
values of the noise distribution stability parameter:
$\alpha=2$ (red triangles), $1.5$ (green diamonds), $1.25$ (blue squares)
and $1$ (black circles).\label{fig:arfima-unbounded}}
\end{figure}

Seeking to generate comparable time series with analyzed in
\cite{Gontis2021Arxiv} order disbalance time series of the financial markets
we choose $N=7 \cdot 10^6$, nine values of $d=\{-0.4,-0.3,-0.2,$
$-0.1,0.0,0.1,0.2,0.3,0.4\}$ and four values of $\alpha=\{2,1.5,1.25,1.0\}$.
The sample time series for any set of parameters have been evaluated using
four estimators described above: MSD, AVE, Higuchi's estimator, and BDA. We
evaluate $H$ as described in the previous subsection. First of all, we
partition time series $Y_i$ in subsets with $5 \cdot 10^5$ time steps and
accumulate them to get $14$ subseries $X_i$. Then the exponent $\lambda$ or
the Hurst parameter are evaluated for each subseries using MSD, AVE, and
Higuchi's sample estimators. Finally, we calculate the mean and standard
deviation of defined $14$ $\lambda$ and $H$ sets. Estimated $d$ we calculate
using $d=H-1/\alpha$ or $d=(\lambda-1)/2$ in MSD case. The graphs in Fig.
\ref{fig:arfima-unbounded} of estimated $d$ versus used ARFIMA model d serve
as a good test of used estimators.

Our numerical result given in subfigure (a) confirms theoretical prediction for the sample MSD $M_N(k) \sim k^{2d+1}$ \cite{Burnecki2010PRE} as estimated $d$ using this relation almost coincide with model $d$ for all values of $\alpha$. It is accepted that two estimators, Absolute value and Higuchi's, are almost
equivalent and should be applicable for the analysis of fractional processes
with stable distribution
\cite{Taqqu1995Fractals,Mercik2003APP,Weron2005PRE,Magdziarz2013JPhysA}.
Indeed, the results of our numerical investigation, see (b) and (c)
subfigures in Fig. \ref{fig:arfima-unbounded} confirm the equivalence of these
estimators. Nevertheless, the estimated values of memory parameter $d$
deviate considerably from its model value, when $\alpha\rightarrow 1$, and
these deviations are much more prominent for the super--diffusion case $d >
0$. These deviations do not arise as a computational effect as estimated relative standard deviation decreases from $0.15$ to $0.02$ for the evaluated $H$ in the investigated interval of $d$. Fortunately, this result does not contradict the study
\cite{Gontis2021Arxiv}, where we used these estimators to evaluate $d$ in
empirical order disbalance time series exhibiting sub--diffusion.

\begin{figure}
\centering
\includegraphics[width=0.7\textwidth]{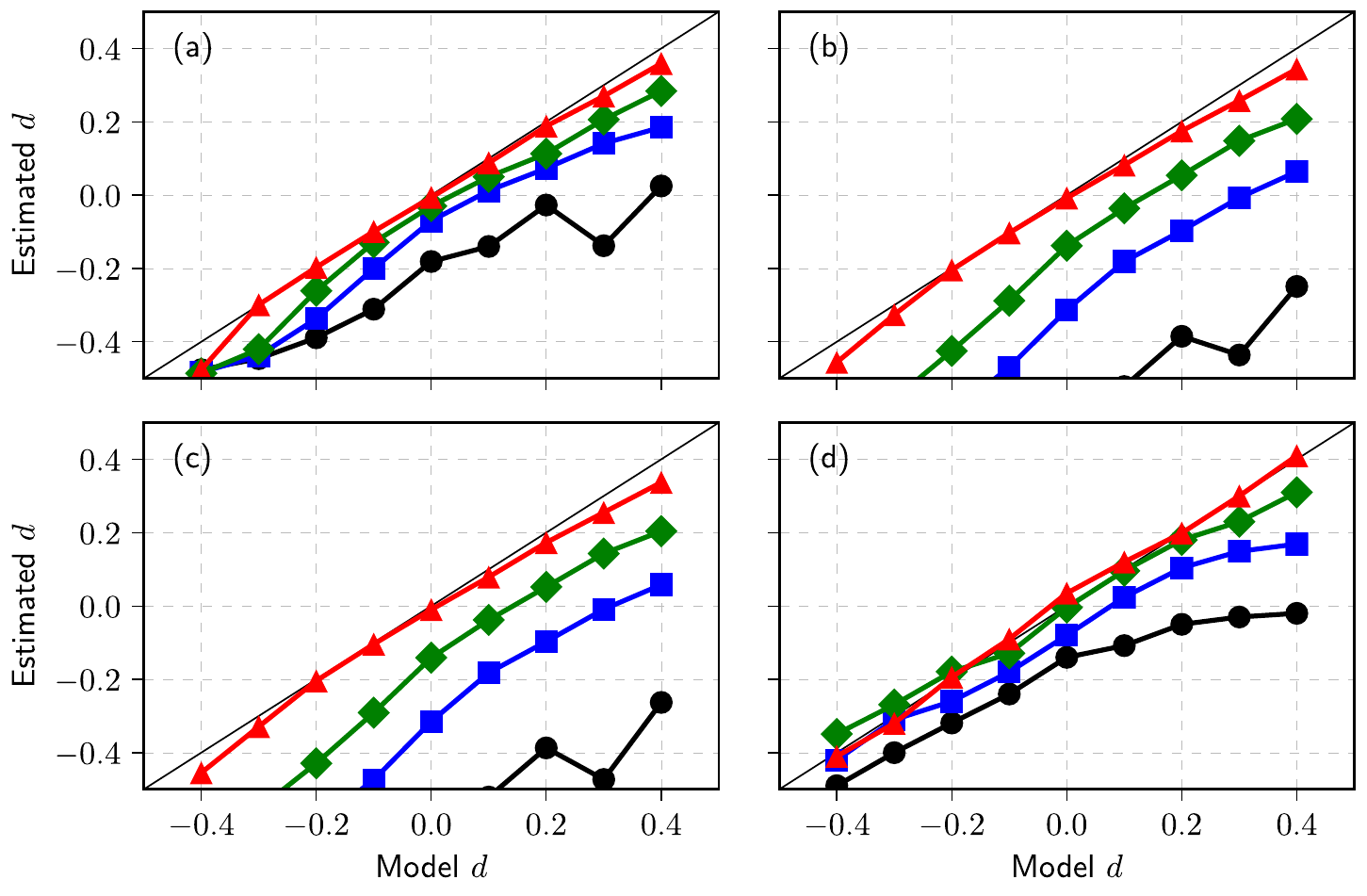}
\caption{Comparison of the MSD (a), AVE (b), Higuchi (c) and BDA
(d) estimator performance when estimating $d$ from the accumulated ARFIMA(0,d,0)
series in the bounded case, $\{X_i\}$ generated by Eq.~\eqref{eq:ARFIMA-acc-bounded}.
Different curves correspond to the different
values of the noise distribution stability parameter:
$\alpha=2$ (red triangles), $1.5$ (green diamonds), $1.25$ (blue squares)
and $1$ (black circles).\label{fig:arfima-bounded}}
\end{figure}

It is important to note that the estimators: MSD, AVE, and Higuchi's should work well only for the unbounded time series when the most physical systems and processes are of finite size and duration. In all such cases, boundary
effects might become important, and one must choose or propose more reliable
estimators \cite{Magdziarz2013JPhysA}. The BDA considered in our previous
work
\cite{Gontis2017PhysA,Gontis2017Entropy,Gontis2018PhysA,Gontis2020JStat},
probably, can serve as an alternative approach. This method works better for
the bounded time series, where more intersections of series with the
threshold can be expected. Thus in this contribution for the BDA we
restrict the diffusion of $X_i$ to the interval $\left[ -X_{max},
X_{max} \right]$ (in our analysis we use $X_{max}=(10^5)^{2d+1}$).
This restriction is implemented as a soft boundary condition:
\begin{equation}
     X_{i+1} = \max\left( \min\left( X_{i}+Y_{i}, X_{max} \right), -X_{max}
     \right) . \label{eq:ARFIMA-acc-bounded}
\end{equation}
This iterative relation replaces Eq.~\eqref{eq:ARFIMA-acc-unbounded} in the
$\{X_i\}$ series generation algorithm.
We define the PDF of the burs and interburst duration $T_j$  for the whole
set of time steps $N=7 \cdot 10^6$ and the series threshold equal to zero
mean. Note that only in this symmetric case PDF's of burst and interbust
duration coincide.  Seeking to understand how the diffusion restriction
mechanism impacts the results of other estimators, we use the same
restriction mechanism for the $14$ subseries obtained after the partition
procedure. Results of this analysis we present in Fig.
\ref{fig:arfima-bounded}.

Though the used diffusion restriction is relatively soft and changes the direction of movement in the limited number of
trajectories points, results of MSD, AVE, and Higuchi's estimators changed
very considerably,  compare subfigures (a), (b) and (c) with corresponding
results in Fig. \ref{fig:arfima-bounded}. Contrary, the results obtained
using $H$ defined by BDA, see subfigure (d), resembles AVE (b) and Higuchi's
estimator (c) subfigures from unbounded series Fig.
\ref{fig:arfima-bounded}. Further investigation is needed to define the best
methods and sample estimators for evaluating parameters of fractional time
series impacted by various diffusion restrictions. The vast amount of data
available from the financial markets can serve as empirical time series
considered from the perspective of FLSM.

\section{Future considerations\label{sec:future-considerations}}

Here we have reviewed our approaches to modeling the long--range memory
phenomenon and power--law statistics in a variety of complex systems. Our
approach differs from the usual approach taken by mathematicians
in that we have used Markovian models instead of the
non--Markovian alternatives. We were able to reproduce similar behaviors due
to our models being driven by various non--linear dependencies. In the case of
SDEs non--linearity may cause the increments of the stochastic process to be
non--stationary and, by consequence, cause spurious long--range memory
\cite{Bassler2006PhysA,McCauley2007PhysA}. Many models we have
built over the years are not the models of the true long--range memory.
However, the critical question is whether our models
capture the memory as observed in the financial markets and possibly
other socioeconomic complex systems.
Section~\ref{sec:arfima}, which describes our most recent endeavor, hints at
three components that are needed to provide an answer.

The first component is a statistical test, which should distinguish
between the spurious and the true long--range memory. Currently, we are
considering BDA method
\cite{Gontis2012ACS,Gontis2017Entropy,Kononovicius2019BDJStat,Gontis2020PhysA},
which performs reasonably well in comparison to the alternatives.
The core idea of the method is that for any
one--dimensional Markovian random walk first--passage time PDF
should be a power--law with exponent $-3/2$ at least for some of the
durations. Deviations from this law could indicate the presence of the true
long--range memory. Though the method may fail when the stochastic process is
not one--dimensional, the study of what happens in the multidimensional case,
e.g., as in \cite{Ruseckas2016JStatMech}, is pending. Other challenges may
also arise, as discussed in Section~\ref{sec:arfima}.

The second component would be a selection of models exhibiting both spurious and
true long--range memory. Our prior research has introduced a variety of models
of spurious long--range memory; hence the next steps would be formulating
comparable alternative models and studying properties of the existing
long--range memory models. Here we have focused on estimating long--range memory in the fractional L\'evy stable motion (modeled
using ARFIMA(0,d,0) discrete process), which is a
generalization of the fractional Brownian motion. However, in general, other
models could also be considered, for example, the multiplicative point
process (see Section~\ref{sec:pp-sde}) could be generalized by replacing
uncorrelated Gaussian noise with fractional Gaussian noise. Other
correlation structures or variable pulse durations could also be considered
as an extension \cite{Ruseckas2003LFZ}. Other notable alternatives and
extensions include continuous--time random walk \cite{Kutner2017EPJB} and
complex contagion frameworks \cite{Baronchelli2018RSOS,Landry2020Chaos}.

The third component would be a variety of data from socioeconomic complex
systems. Many of our earlier approaches relied on high--frequency absolute
return and trading activity time series, but in our most recent works, we
have shifted our attention to the order book data obtained from LOBSTER
\cite{Huang2011Lobster}. Order book data seems to invite a more general
approach by understanding the data within FLSM or ARFIMA mindset for a broad class of anomalous diffusion processes.
\cite{Weron2005PRE,Magdziarz2013JPhysA,Burnecki2014JStatMech}. The vast
amount of data in social including financial systems, has to be investigated
to identify and validate the fractional dynamics and long--range memory. Our
first results in this direction \cite{Gontis2020JStat,Gontis2021Arxiv}
question the interpretation of long--range memory in the order flow data of
financial markets. First of all, a prudent choice of estimators based on
FLSM and ARFIMA assumptions are needed. After extensive analysis from this
perspective would be possible to decide whether the investigated social system
exhibits true long--range memory or observed power--law statistical properties
are just the outcome of strong non--linear effects.

Research effort combining all these three components could yield a better
understanding of the long--range memory phenomenon as it is observed in the
variety of complex systems.

The comprehensive interpretation of long-range memory observed in the
financial and other social systems should considerably contribute to
developing advanced analytical tools for applications in financial markets.
Thus, we have focused on the description and explanation of the long--range
memory phenomenon.  Notably, a few more recent works refer to or use some of
our results and are more application-minded. In \cite{Lera2015} a
non--linear SDE was derived, providing both physical and economic arguments,
to study the performance of EUR/CHF exchange rate. The derived SDE belongs
to the class described by \eqref{eq:sde-class}.  \cite{Leibovich2016PRE} has
considered the relationship between aging and long--range memory phenomena
in a couple of physics experiments: blinking-quantum-dots, single-file
diffusion, and Brownian motion in a logarithmic potential.
\cite{Dmitriev2017PCS} has shown that SDE~\eqref{eq:sde-class} applies to
the modeling of the dynamics on microblogging networks.
\cite{DiVita2019EPJB} has considered the effects of perturbations on the
stability of power--law distributions in general with an application to
wealth distributions. \cite{Ponta2019PhysA} tested the applicability of
simple stochastic models to the modeling of non--stationary behavior of
intraday tick--by--tick returns.  \cite{Emenogu2020} has tested forecast
robustness of nonlinear GARCH model when time series exhibit high positive
autocorrelation. Mean reversion phenomenon was studied in Karachi Stock
Exchange data from the perspective of GARCH models in \cite{Vveinhard2016}.
\cite{Lima2021PhysA} has compared the performance of non--linear SDE models
against Black and Scholes model, which is one of the models used by the
practitioners. Various modifications of Heston model, another model favored
by the practitioners, are also reminiscent of SDE~\eqref{eq:sde-class}
\cite{Benhamou2010SIAM}. In the future we hope to inspire and maybe take up
more application--minded endeavors.

{\small
\setstretch{1.0}
\section*{Author contributions}

Conceptualization, R.K., A.K., B.K and V.G.; methodology, R.K., A.K., B.K
and V.G.; software,  A.K. and V.G.; validation, R.K., A.K., B.K and V.G.;
formal analysis, B.K.; investigation, R.K., A.K., B.K and V.G.; resources,
R.K., A.K., B.K and V.G.; data curation, R.K., A.K., B.K and V.G.;
writing---original draft preparation, R.K., A.K., B.K and V.G.;
writing---review and editing, R.K.; visualization, R.K., A.K., B.K and V.G.;
supervision, V.G.; project administration, R.K.; funding acquisition, R.K.
and V.G.

\section*{Funding}

This project was funded by the European Union (project No
09.3.3-LMT-K-712-19-0017) under the agreement with the Research Council of
Lithuania (LMTLT).

\section*{Abbreviations}

\begin{tabular}{@{}ll}
ABM & agent--based model\\
ARCH & autoregressive conditional heteroscedasticity\\
ARFIMA & autoregressive fractionally integrated moving average\\
AVE & absolute value estimator\\
BDA & burst and interburst duration analysis\\
FBM & fractional Brownian motion\\
FGN & fractional Gaussian noise\\
FIGARCH & fractionally integrated GARCH\\
FLSM & fractional L\`{e}vy stable motion\\
FLSN & fractional L\`{e}vy stable noise\\
GARCH & generalized ARCH\\
MSD & mean squared displacement\\
NYSE & New York stock exchange\\
PDF & probability density function\\
PSD & power spectral density \\
SDE & stochastic differential equation \\
VSE & Vilnius stock exchange \\
WSE & Warsaw stock exchange
\end{tabular}%
}

\bibliographystyle{elsarticle-num}
\bibliography{manuscript}

\begingroup
\small
\setstretch{1.0}

\endgroup

\end{document}